\documentstyle[12pt]{article}

\begin{document}

\begin{flushright}

IMSc/2017/06/04 

\end{flushright} 

\vspace{2mm}

\vspace{2ex}

\begin{center}

{\large \bf Variety of $(d + 1)$ dimensional } \\

\vspace{4ex}

{\large \bf Cosmological Evolutions with and without bounce} \\

\vspace{4ex}

{\large \bf in a class of LQC -- inspired Models} \\

\vspace{8ex}

{\large  S. Kalyana Rama}

\vspace{3ex}

Institute of Mathematical Sciences, HBNI, C. I. T. Campus, 

\vspace{1ex}

Tharamani, CHENNAI 600 113, India. 

\vspace{2ex}

email: krama@imsc.res.in \\ 

\end{center}

\vspace{6ex}

\centerline{ABSTRACT}

\begin{quote} 

The bouncing evolution of an universe in Loop Quantum Cosmolgy
can be described very well by a set of effective equations,
involving a function $sin \; x$. Recently, we have generalised
these effective equations to $(d + 1)$ dimensions and to any
function $f(x) \;$. Depending on $f(x) \;$ in these models
inspired by Loop Quantum Cosmolgy, a variety of cosmological
evolutions are possible, singular as well as non singular. In
this paper, we study them in detail. Among other things, we find
that the scale factor $a(t) \; \propto \; t^{ \frac {2 q} {(2 q
- 1) \; (1 + w) d}} \;$ for $f(x) = x^q \;$, and find explicit
Kasner--type solutions if $w = 2 q - 1 \;$ also. A result which
we find particularly fascinating is that, for $f(x) = \sqrt{x}
\;$, the evolution is non singular and the scale factor $a(t)$
grows exponentially at a rate set, not by a constant density,
but by a quantum parameter related to the area quantum.

\end{quote}

\vspace{2ex}


\vspace{2ex}









\newpage

\vspace{4ex}

\begin{center}

{\bf 1. Introduction} 

\end{center}

\vspace{2ex}

Consider a homogeneous expanding universe whose constituents
have the density $\rho$ and the pressure $p$ obeying the
condition $\rho + p > 0 \;$. In Einstein's theory of general
relativity, as one goes back in time, the size of such an
universe decreases and vanishes at a finite time in the
past. The curvature invariants then diverge, the density
diverges, and the evolution of the universe becomes singular.

In a quantum theory of gravity, among other things, such
singularities are expected to be resolved. In Loop Quantum
Cosmology (LQC) \cite{b01} -- \cite{10shyam} which arises in a
$(3 + 1)$ dimensional quantum gravity theory based on Loop
Quantum Gravity (LQG) \cite{ashtekar, books}, such singularities
are indeed resolved due to quantum effects. In LQC, as one goes
back in time, the size of the universe decreases, reaches a non
zero minimum, and increases to $\infty$ in the infinite
past. The curvature invariants remain finite, the density
remains bounded from above and, therefore, the evolution of the
universe in LQC has a bounce and is non singular.

It turns out that the quantum dynamics of such a non singular
evolution in LQC can be described very well by a set of
effective equations \cite{aps, status, 05shyam, aw, barrau,
singh}. These equations reduce to Einstein's equations in the
classical limit. These effective equations, in our notation,
involve a certain function $f(x) \;$, see equations
(\ref{hgrav}) and (\ref{genpsi}) where it will be first
introduced. The variable $x$ will turn out to be related to the
time derivative of the scale factor, and the function $f$ will
dictate the precise relation between them. In LQC, this function
$f(x) = sin \; x$ and in the classical limit which leads to
Einstein's equations, $f(x) = x \;$. In fact, Einstein's
equations follow upto a scaling of time whenever $f(x)$ vanishes
linearly. Thus, in LQC where $f(x) = sin \; x \;$, the evolution
is same as in Einstein's theory in the limits $x \to 0$ as well
as $x \to \pi \;$. These limits constitute the two ends of the
bouncing evolution in LQC where the size of the universe evolves
to infinity, and the evolution in these two asymptotic limits is
same as in Einstein's theory.

In a recent paper \cite{k16}, we have generalised the effective
LQC equations. Our generalisations are empirical and not derived
from any underlying theory. But they are simple,
straightforward, and natural. We generalised from $(3 + 1)$ to
$(d + 1)$ dimensions where $d \ge 3 \;$, \footnote{ There exists
a $(d + 1)$ dimensional LQG formulation, given in \cite{th1,
th2, th3}. Our preliminary analysis \cite{work}, see
\cite{zhang} also, suggests that one can derive the LQC analogs
of the effective equations in $(d + 1)$ dimensions, with $f(x) =
sin \; x \;$.} and generalised the trigonometric function as
well as a $\bar{\mu}$ function which appear in the effective LQC
equations. These generalised equations describe the cosmological
evolution of a $(d + 1)$ dimensional homogeneous, anisotropic
universe and may be considered as a class of LQC -- inspired
models.

In this paper, we will consider only the generalisation of the
trigonometric function, keeping the $\bar{\mu}$ function similar
to that in the so-called $\bar{\mu}-$scheme. In LQC, the
function $f(x) = sin \; x$ is determined by the underlying
theory. In the present LQC -- inspired models, the
generalisation is empirical and no underlying theory is invoked
which may determine $f(x) \;$. This feature is a shortcoming of
the present models. But taking it as a strength, one may
consider a variety of functions $f(x)$ from a completely general
point of view, study the corresponding evolution, and gain
insights into the various types of singular and non singular
evolutions possible.

Here, we follow this approach and study a variety of $(d + 1)$
dimensional cosmological evolutions in the LQC -- inspired
models, corresponding to a variety of possible behaviours of the
function $f(x) \;$. Assuming that $p = w \; \rho$ where $w$ is a
constant and $(1 + w) > 0 \;$, we study both isotropic and
anisotropic cases. The resulting evolutions may or may not have
a bounce, and may be singular or non singular. In the isotropic
case, for several functions $f(x) \;$, we also find the
potential $V(\sigma)$ for a minimally coupled scalar field
$\sigma$ which may give rise to the equation of state $p = w \;
\rho \;$. See \cite{Vlq} for analogous potential $V(\sigma)$ in
LQC. 

In general, given a function $f(x) \;$, it is not possible to
obtain explicit solutions to the relevant equations of motion.
However, these equations have a shift and a scaling symmetry
which may be used to understand several important features of
the evolutions. In this paper, we first consider the cases with
non trivial functions $f(x)$ where explicit solutions may be
obtained. For the isotropic case, explicit solutions may be
obtained for the functions $f(x) = sin \; x \; , \; x^q \;$, and
$e^x \;$. For the anisotropic case, explicit solutions may be
obtained for the function $f(x) = x^q \;$ if $w = 2 q - 1 \;$.
These anisotropic solutions are the analogs of the standard
Kasner--type solutions in Einstein's theory.

We then study a class of functions where $f(x) \to x$ as $x \to
0$ so that the evolution in this limit is same as in Einstein's
theory. Thus, in this limit, the time $t \to \infty \;$, the
scale factor $a \to \infty \;$, and the universe is expanding.
As $x$ increases, $t$ decreases, $a$ decreases, and the universe
decreases in size. Assuming that $f(x)$ remains positive and all
its derivatives remain bounded for $0 < x < x_r \;$, we study
possible asymptotic behaviours in cases where $f(x) \to 0 \;$ or
$\infty \;$ in the limit $x \to x_r \;$, and $x_r$ itself may be
finite or infinite. We study the cases where the function $f(x)
\propto (x_r - x)^q $ in the limit $x \to x_r \;$ and also the
cases where, in the limit $x \to \infty \;$, the function $f(x)
\propto x^q \;$, or $f(x) \to (const) \;$, or $f(x) \propto e^{-
b \; x} \;$. Such asymptotic behaviours are quite natural and,
hence, they may apply to a wide class of functions $f(x) \;$.
Together with the shift and the scaling symmetries, the explcit
solutions obtained earlier may now be used to describe the
asymptotic evolutions in all of these cases.

The main results of our study are the following. The explicit
solution presented in this paper for $f(x) = sin \; x$
generalises the $(3 + 1)$ dimensional LQC solution given in
\cite{Vlq} to $(d + 1)$ dimensions. The explicit solutions
presented in this paper for $f(x) = x^q $ generalise the
evolution of the scale factor $a$ in Einstein's theory to $a(t)
\; \propto \; t^{ \frac {2 q} {(2 q - 1) \; (1 + w) d}} \;$.
When $w = 2 q - 1 \;$, they also generalise the standard
Kasner--type solutions in Einstein's theory.

We find both singular and non singular evolutions. The
asymptotic behaviours of the scale factor in the isotropic cases
are tabulated in Tables I and II. Here, in the Introduction, we
point out a few cases of non singular evolutions. In all these
cases, in the limit $x \to 0 \;$, the function $f(x) \to x \;$,
the scale factor $a \to \infty \;$, the time $t \to \infty \;$,
and $a$ and $t$ decrease as $x$ increases.

\begin{itemize}

\item

For the case where $f(x) \propto (x_r - x)^q \;$ as $x \to x_r
\;$ and $2 q \ge 1 \;$, the evolution is non singular and has a
bounce. In the limit $x \to x_r \;$, one has $t \to - \infty \;$
and $a \to \infty \;$. The function $f(x) = sin \; x$ falls
under this case and corresponds to $x_r = \pi$ and $q = 1 \;$.

\item

For the case where $f(x) \propto x^q \;$ as $x \to \infty \;$
and $2 q \le 1 \;$, the evolution is non singular. In the limit
$x \to \infty \;$, one has $t \to - \infty \;$; and $a \to 0 \;$
if $0 < 2 q \le 1 \;$, $\; a \to (const) \;$ if $q = 0 \;$, and
$a \to \infty \;$ if $q < 0 \;$. The evolution for the $q = 0$
case in this limit is similar to that expected in the Hagedorn
phase of string/M theory \cite{bowick} -- \cite{k06}.

\end{itemize}

We find that the case where $f(x) = x^q$ and $0 < 2 q < 1$
exhibit an interesting feature : There is a singularity when the
universe is increasing to infinite size, and no singularity when
it is contracting to zero size. This is opposite to what one
usually comes across in Einstein's theory.

We find that the $2 q = 1 \;$ case where $f(x) = \sqrt{x}$ is
particularly fascinating. The evolution now is non singular and
it straddles the border between the singular and non singular
evolution : in the limit $x \to 0 \;$, the evolution is non
singular for $2 q > 1$ and is singular for $2 q < 1 \;$; in the
limit $x \to \infty \;$, the evolution is singular for $2 q > 1$
and is non singular for $2 q < 1 \;$.

Also, for $2 q = 1 \;$, the scale factor $a$ grows exponentially
and the exponential rate is set, not by a constant density, but
by a quantum parameter which is related to the area quantum as
in LQC. The density $\rho \propto a^{- (1 + w) \; d} \;$ and is
not constant. This exponential growth of the scale factor is,
therefore, unlike that which occurs in Einstein's theory due to
a positive cosmological constant for which $w = - 1 \;$ and the
density is constant. We find these results intriguing and
fascinating but their significance, if any, is not clear to us
at present.

This paper is organised as follows. In Section {\bf 2}, we
present the equations of motion for a $(d + 1)$ dimensional
homogeneous anisotropic universe in Einstein's theory. In
Section {\bf 3}, we present the effective equations of motion in
LQC. In Section {\bf 4}, we present the generalised effective
equations of motion for our LQC -- inspired models. In Section
{\bf 5}, we describe the general features of these models. In
Section {\bf 6}, we present explicit isotropic solutions and use
them to study other asymptotic evolutions. In Tables I and II,
we have also tabulated the asymptotic behaviours of the scale
factor in the isotropic cases so that the results may be seen at
a glance. In Section {\bf 7}, we present explicit anisotropic
solutions and study other anisotropic asymptotic evolutions. In
Section {\bf 8}, we present a brief summary and conclude by
mentioning several issues for further studies.


\vspace{4ex}

\begin{center}

{\bf 2. $(d + 1)$ dimensional Einstein's equations}

\end{center}

\vspace{2ex}

Let the spacetime be $(d + 1)$ dimensional where $d \ge 3 \;$
and let $x^i$, $\; i = 1, 2, \cdots, d$, denote the spatial
coordinates. Also, let the $d-$dimensional space be toroidal and
let $L_i$ denote the coordinate length of the $i^{th}$
direction. Consider a homogeneous and anisotropic universe whose
line element $d s$ is given by
\begin{equation}\label{ds}
d s^2 = - d t^2 + \sum_i a_i^2 \; (d x^i)^2 
\end{equation}
where the scale factors $a_i \;$ depend on $t$ only. Here and in
the following, we will explicitly write the indices to be summed
over since the convention of summing over repeated indices is
not always applicable. Einstein's equations are given, in the
standard notation with $\kappa^2 = 8 \pi G_{d + 1} \;$, by
\begin{equation}\label{rab}
R_{A B} - \frac{1}{2} \; g_{A B} R = \kappa^2 \; T_{A B}
\; \; \; , \; \; \; \;
\sum_A \nabla^A T_{A B} = 0 
\end{equation}
where $A, B = (0, i)$ and $T_{A B}$ is the energy momentum
tensor. We assume that $T_{A B}$ is diagonal and that its
diagonal elements are given by
\begin{equation}\label{0i}
T_{0 0} = \rho \; \; , \; \; \; T_{i i} = p_i
\end{equation}
where $\rho$ is the density and $p_i$ is the pressure in the
$i^{th}$ direction. Defining the quantities $\lambda^i, \;
\Lambda, \; a, \; G_{i j}$, and $G^{i j}$ by
\[
e^{\lambda^i} = a_i \; \; , \; \; \; 
e^\Lambda = \prod_i a_i = a^d 
\; \; \; \longrightarrow \; \; \;
\Lambda = \sum_i \lambda^i \; \; , 
\]
\[
G_{i j} = 1 - \delta_{i j} \; \; , \; \; \; 
\sum_i G^{i j} G_{j k} = \delta^i_{\; k}
\; \; \; \longrightarrow \; \; \;
G^{i j} = \frac{1}{d - 1} - \delta^{i j} \; \; , 
\]
and after a straightforward algebra, Einstein's equations
(\ref{rab}) give
\begin{eqnarray}
\sum_{i j} G_{i j} \; \lambda^i_t \; \lambda^j_t
& = & 2 \kappa^2 \; \rho \label{t21} \\
& & \nonumber \\
\lambda^i_{t t} + \Lambda_t \; \lambda^i_t & = & \kappa^2 \;
\sum_j G^{i j} \; (\rho - p_j) 
\label{t22} \\
& & \nonumber \\
\rho_t + \sum_i (\rho + p_i) \; \lambda^i_t & = & 0 \label{t23}
\end{eqnarray}
where the $t-$subscripts denote derivatives with respect to $t
\;$. It follows from equations (\ref{t22}) that
\begin{equation}\label{t22a}
\left( \lambda^i_t - \lambda^j_t \right)_t
+ \Lambda_t \; \left( \lambda^i_t - \lambda^j_t \right)
\; = \; \kappa^2 \; (p_i - p_j) \; \; .
\end{equation}
If the pressures are isotropic then $p_i = p$ for all $i \;$,
and we get
\begin{equation}\label{t22b}
\lambda^i_t - \lambda^j_t = (const) \; e^{- \Lambda} 
\end{equation}
and
\begin{equation}\label{tiso3}
\rho_t + \Lambda_t \; (\rho + p) = 0 \; \; . 
\end{equation}
If the scale factors are also isotropic then $a_i = a$ and
$\lambda^i_t = \frac {a_t} {a} \equiv H \;$ for all $i \;$, and
equations (\ref{t21}) and (\ref{t22}) give 
\begin{eqnarray}
H^2 & = & \frac {2 \; \kappa^2 \; \rho} {d (d - 1)}
\label{tiso1} \\
& & \nonumber \\
H_t & = & - \; \frac {\kappa^2 \; (\rho + p)} {d - 1}
\label{tiso2} \; \; .
\end{eqnarray}

Note that if $p_i = w_i \; \rho$ where $w_i$ are constants then
equations (\ref{t21}) -- (\ref{t23}) can be solved exactly
\cite{k16}. Also note that for a minimally coupled scalar field
$\sigma$, which has a potential $V(\sigma) \;$ and depends on
$t$ only, it is a standard result that the density $\rho$ and
the pressure $p$ are given by
\begin{equation}\label{rhopsigma}
\rho = \frac {(\sigma_t)^2} {2} + V \; \; , \; \; \;
p_i = p  = \frac {(\sigma_t)^2} {2} - V \; \; .
\end{equation}
The equation of motion for the field $\sigma$ is given by
\begin{equation}\label{sigmaeom}
\sigma_{t t} + \Lambda_t \; \sigma_t + \frac{d V}{d \sigma} = 0
\end{equation}
which also follows from equations (\ref{tiso3}) and
(\ref{rhopsigma}). Furthermore, in the isotropic case, one can
construct a potential $V$ for the scalar field $\sigma$ such
that $p = w \rho$ where $w$ is a constant: Writing $\sigma_t^2 =
(1 + w) \; \rho$ and $2 V = (1 - w) \; \rho$ and after some
manipulations involving equations (\ref{tiso1}) and
(\ref{tiso2}), see comments below equation (\ref{v2}), it can
be shown that the required potential is given by
\begin{equation}\label{v1}
V(\sigma) \propto (1 - w) \; e^{- 2 c_w \sigma}
\; \; , \; \; \;
c_w = \sqrt{ \frac {(1 + w) \; \kappa^2 d} {2 \; (d - 1)}} \; \; . 
\end{equation}


\vspace{4ex}

\begin{center}

{\bf 3. $(3 + 1)$ dimensional Effective LQC equations}

\end{center}

\vspace{2ex}

In this section, we mention briefly the main steps involved in
obtaining the effective equations of motion in the $(3 + 1)$
dimensional Loop Quantum Cosmology (LQC). A detailed derivation
and a complete description of various terms and concepts
mentioned below are given in the review \cite{status}. Below, we
present the LQC expressions in a form which can be readily
generalised.

Let the three dimensional space be toroidal and let the line
element $d s$ be given by equation (\ref{ds}) where $d = 3$ now,
and $i = 1, 2, 3 \;$. Let $L_i$ and $a_i L_i$ be the coordinate
and the physical lengths of the $i^{th}$ direction. In the Loop
Quantum Gravity (LQG) formalism, the canonical pairs of phase
space variables consist of an $SU(2)$ connection $A^i_a =
\Gamma^i_a + \gamma K^i_a$ and a triad $E^a_i$ of density weight
one. Here $\Gamma^i_a$ is the spin connection defined by the
triad $e^a_i$, $\; K^i_a$ is related to the extrinsic curvature,
and $\gamma > 0$ and $\approx 0.2375$ is the Barbero -- Immirzi
parameter of LQG, its numerical value being suggested by the
black hole entropy calculations. For the anisotropic universe,
whose line element $d s$ is given in equation (\ref{ds}) with $d
= 3$, one has $A^i_a \propto \hat{c}_i$ and $E^a_i \propto
\hat{p}_i$ where $\hat{c}_i \;$ will turn out to be related to
the time derivative of $a_i$, and $\hat{p}_i$ is given by
\begin{equation}\label{pi}
\hat{p}_i = \frac {V} {a_i L_i} \; \; , \; \; \;
V = \prod_j {a_j L_j}
\end{equation}
with $V$ being the physical volume. The full expressions for
$A^i_a$ and $E^a_i$ contain various fiducial triads, cotriads,
and other elements, and are given in \cite{status, aw}. The non
vanishing Poisson brackets among $\hat{c}_i$ and $\hat{p}_j$ are
given by
\begin{equation}\label{cipi}
\{ \hat{c}_i, \; \hat{p}_j \} = \gamma \; \kappa^2 \;
\delta_{i j}
\end{equation}
where $\kappa^2 = 8 \pi G_4 \;$. The effective equations of
motion are given by the `Hamiltonian constraint' ${\cal C}_H =
0$ and by the Poisson brackets of $\hat{p}_i$ and $\hat{c}_i$
with ${\cal C}_H$ which give the time evolutions of $\hat{c}_i$
and $\hat{p}_i \;$: namely, by
\begin{equation}\label{dynamics} {\cal C}_H = 0 \; \; , \; \; \;
(\hat{p}_i)_t = \{ \hat{p}_i, \; {\cal C}_H \} \; \; , \; \; \;
(\hat{c}_i)_t = \{ \hat{c}_i, \; {\cal C}_H \} \; \; . 
\end{equation} 

Given that Einstein's action for gravity is known, it is to be
expected that there exists a classical ${\cal C}_H \;$, the
Poisson brackets with which lead to the classical dynamics given
by Einstein's equations. Non trivially, and as reviewed in
detail in \cite{status}, there also exists an effective ${\cal
C}_H \;$, the Poisson brackets with which lead to the equations
of motion which describe very well the quantum dynamics of
LQC. The effective ${\cal C}_H$ reduces to the classical one in
a suitable limit.

The expression for the ${\cal C}_H \;$ is of the form
\begin{equation}\label{chtot}
{\cal C}_H = H_{grav} (\hat{p}_i, \; \hat{c}_i) + H_{mat}
(\hat{p}_i \; ; \; \{ \phi_{mat} \}, \; \{ \pi_{mat} \}) 
\end{equation}
where $H_{grav}$ denotes the effective gravitational Hamiltonian
and $H_{mat}$ denotes a generalised matter Hamiltonian. In the
matter sector, the density $\rho$ and the pressure $p_i$ in the
$i^{th}$ direction are defined by
\begin{equation}\label{phati}
\rho = \frac {H_{mat}} {V} \; \; , \; \; \;
p_i = - \; \frac {a_i L_i} {V} \;
\frac {\partial H_{mat}} {\partial (a_i L_i)} \; \; .
\end{equation}
The pressure $p_i$ is thus, as to be physically expected,
proportional to the change in energy per fractional change in
the physical length in the $i^{th}$ direction. As indicated in
equation (\ref{chtot}), $\; H_{mat}$ is assumed to be
independent of $\hat{c}_i \;$. Since $\hat{c}_i$ will turn out
to be related to $(a_i)_t \;$, this assumption is equivalent to
assuming that matter fields couple to the metric fields but not
to the curvatures. This assumption can also be shown to lead to
the conservation equation (\ref{t23}), namely to 
\begin{equation}\label{ceqn}
\rho_t = \left( \frac {H_{mat}} {V} \right)_t
= - \; \sum_i (\rho + p_i) \; \lambda^i_t \; \; , 
\end{equation}
irrespective of what $H_{grav}$ is \cite{k16}.

In the gravitational sector, the effective $H_{grav}$, from
which the LQC dynamics follow, is given in the so--called
$\bar{\mu}-$scheme by
\begin{equation}\label{hgravlqc}
H_{grav} = - \; \frac {V} {\gamma^2 \lambda_{qm}^2 \kappa^2}
\; \left( sin (\bar{\mu}^1 \hat{c}_1) \; sin (\bar{\mu}^2
\hat{c}_2) + \; cyclic \; \; terms \right)
\end{equation}
where $V = \sqrt{\hat{p}_1 \hat{p}_2 \hat{p}_3}$ is the physical
volume, $\lambda_{qm}^2 = \sqrt{ \frac {3} {4}} \; \gamma
\kappa^2$ is the quantum of area, and $\bar{\mu}^i = \frac
{\lambda_{qm} \hat{p}_i} {V}$ in the $\bar{\mu}-$scheme.
Classical $H_{grav}$ follows in the limit $\bar{\mu}^i \hat{c}_i
\to 0 \;$ where $sin \; (\bar{\mu}^i \hat{c}_i) \to \bar{\mu}^i
\hat{c}_i \;$.

\newpage

\vspace{4ex}

\begin{center}

{\bf 4. $(d + 1)$ dimensional LQC -- inspired models}

\end{center}

\vspace{2ex}

In a recent paper \cite{k16}, we generalised the effective LQC
equations. Our generalisations are empirical and not derived
from any underlying theory. But they are simple,
straightforward, and natural. And, they may be used to model a
variety of non singular cosmological evolutions. In \cite{k16},
we generalised from $(3 + 1)$ to $(d + 1)$ dimensions where $d
\ge 3 \;$, and generalised the trigonometric and the $\bar{\mu}$
functions appearing in the effective $H_{grav}$ in equation
(\ref{hgravlqc}). In this paper, we will consider only the
generalisation of the trigonometric function, keeping the
$\bar{\mu}$ function as in the $\bar{\mu}-$scheme.

We now present the generalised effective equations of our LQC --
inspired models, expressing them so that they resemble equations
(\ref{t21}) and (\ref{t22}) as closely as possible. For the
purpose of this generalisation, we have already presented the
LQC expressions in a form which can be readily taken over. Upon
generalisation, we have the following.

\begin{itemize}

\item


The index $i = 1, 2, \cdots, d \;$ now in the LQC expressions.

\item


The canonical pairs of phase space variables are given by
$\hat{c}_i$ which will be related to $(a_i)_t$, and $\hat{p}_i$
which is given by equation (\ref{pi}). The non vanishing Poisson
brackets among $\hat{c}_i \;$ and $\hat{p}_j$ are given by
equation (\ref{cipi}) where now $\kappa^2 = 8 \pi G_{d + 1} \;$
and $\gamma$ may characterise the quantum of the $(d - 1)$
dimensional area given by $\lambda_{qm}^{d - 1} \sim \gamma
\kappa^2$ \cite{th1, th2, th3, nb}.

\item


The effective equations of motion are given by equation
(\ref{dynamics}) where ${\cal C}_H \;$ is of the form given in
equation (\ref{chtot}). In the matter sector, the density $\rho$
and the pressures $p_i$ are given by equations (\ref{phati}),
and they satisfy the standard conservation equation
(\ref{ceqn}).

\item


In the gravitational sector, the effective $H_{grav}$ in
equation (\ref{hgravlqc}) is now generalised to
\begin{equation}\label{hgrav}
H_{grav} = - \; \frac {V \; {\cal G}} {\gamma^2 \lambda_{qm}^2
\kappa^2} \; \; , \; \; \; \; {\cal G} = \frac{1}{2}
\sum_{i j} G_{i j} f^i f^j = \sum_{i j \; (i < j)} f^i f^j
\end{equation}
where $V = \left( \prod_i \hat{p}_i \right)^{\frac{1}{d - 1}}$
is the $d$ dimensional physical volume and
\begin{equation}\label{genpsi} 
f^i = f(m^i) \; \; , \; \; \;  
m^i = \bar{\mu}^i \; \hat{c}_i
\; \; , \; \; \;  
\bar{\mu}^i = \frac {\lambda_{qm} \; \hat{p}_i} {V} \; \; .
\end{equation}
The function $f(x)$ which appears in equation (\ref{genpsi}) is
arbitrary, but with the only requirement that $f(x) \to x$ as $x
\to 0 \;$ so that classical $H_{grav}$ is obtained in the limit
$m^i \to 0 \;$. Clearly, LQC is obtained upon setting $d = 3$
and $f(x) = sin \; x \;$.

\end{itemize}


\vspace{4ex}

\begin{center}

{\bf Equations of motion}

\end{center}

\vspace{2ex}

Equations of motion can now be obtained using the generalised
$H_{grav}$ given in equation (\ref{hgrav}). They will describe
the evolution of a $(d + 1)$ dimensional homogeneous anisotropic
universe in our LQC -- inspired models. The required algebra is
straightforward but involved, see \cite{k16} for details. In
this paper, we present only the final equations which suffice
for our purposes here. The resulting equations of motion,
expressed so that they resemble equations (\ref{t21}) and
(\ref{t22}) as closely as possible, are given by
\begin{eqnarray}
\sum_{i j} G_{i j} f^i f^j & = & 2 \; \gamma^2 \lambda_{qm}^2
\kappa^2 \; \rho \label{e1} \\
& & \nonumber \\
(m^i)_t \; + \; \sum_j \frac {(m^i - m^j) \; X_j} {(d - 1) \;
\gamma \lambda_{qm}} & = & - \; \gamma \lambda_{qm} \kappa^2 \;
\sum_j G^{i j} \; (\rho + p_j) \label{e2} \\
& & \nonumber \\ 
(\gamma \lambda_{qm}) \; \lambda^i_t & = & \sum_j G^{i j} X_j
\label{e3} 
\end{eqnarray}
where we have defined 
\begin{equation}\label{xigi}
X_i = g_i \sum_j G_{i j} f^j \; \; , \; \; \;
g_i = \frac{d f(m^i)} {d m^i} \; \; . 
\end{equation}
Equations (\ref{e1}) -- (\ref{e3}) give the conservation
equation (\ref{t23}). Also, equation (\ref{e3}) gives $(d - 1)
\; (\gamma \lambda_{qm}) \; \Lambda_t = \sum_j X_j \;$, and
equation (\ref{e2}) then gives
\begin{equation}\label{e4}
(m^i - m^j)_t + \Lambda_t \; (m^i - m^j) \; = \; \gamma
\lambda_{qm} \kappa^2 \; (p_i - p_j) \; \; .
\end{equation}
If the pressures are isotropic then $p_i = p$ for all $i \;$,
and we get
\begin{equation}\label{e4b}
m^i - m^j = \mu^{i j} \; e^{- \Lambda} 
\end{equation}
where $\mu^{i j}$ are constants. In the completely isotropic
case, we have
\begin{equation}\label{iso}
(p_i , \; m^i , \; f^i , \; a_i ) = (p , \; m , \; f , \; a )
\end{equation}
and, hence, 
\[
\lambda^i_t = \frac {a_t} {a} = H \; \; , \; \; \; 
X_i = (d - 1) \; g f \; \; , \; \; \; 
g_i = g = \frac{d f} {d m} \; \; .
\]
Equations (\ref{e1}) -- (\ref{e3}) then give
\begin{eqnarray}
f^2 & = & \frac {2 \; \gamma^2 \lambda_{qm}^2 \kappa^2 \;
\rho} {d \; (d - 1)} \; = \; \frac {\rho} {\rho_{qm}}
\label{eiso1} \\
& & \nonumber \\
m_t & = & - \; \frac {\gamma \lambda_{qm} \kappa^2} {d - 1} \;
(\rho + p) \label{eiso2} \\
& & \nonumber \\ 
H & = & \frac {g \; f} {\gamma \lambda_{qm}} \label{eiso3} \\
& & \nonumber 
\end{eqnarray}
where $\rho_{qm} = \frac {d \; (d - 1)} {2 \; \gamma^2
\lambda_{qm}^2 \kappa^2} \;$. Equations (\ref{eiso1}) --
(\ref{eiso3}) give the conservation equation (\ref{tiso3}). We
will assume that $\rho + p > 0 \;$. Then equation (\ref{eiso2})
gives $m_t < 0 \;$. Hence, $m$ will increase monotonically as
$t$ decreases. Also, equations (\ref{eiso1}) and (\ref{eiso3})
give
\begin{equation}\label{eiso3a}
H^2 \; = \; \frac {2 \; \kappa^2 \; (\rho \; g^2)} 
{d \; (d - 1)} \; \; .
\end{equation}
If $g$ can be expressed in terms of $f$ then, using $f^2 = 
\frac {\rho} {\rho_{qm}} \;$, one can express $H^2$ in terms of
$\rho \;$ alone. For example, 
\begin{equation}\label{flqc}
f(x) = sin \; x \; \; \; \Longrightarrow \; \; \;
H^2 \; \propto \; \rho \; \left( 1 - \frac {\rho} {\rho_{qm}}
\right)
\end{equation}
and $f(x) = f(x \; ; \; n)$ where $n$ is a positive integer and 
\begin{equation}\label{fn}
f(x \; ; \; n) = 1 - \left( 1 - \frac {x} {x_*}
\right)^{2 n} \; \; \; \Longrightarrow \; \; \;
H^2 \; \propto \; \rho \; \left( 1 - \sqrt{ \frac {\rho}
{\rho_{qm}}} \; \right)^{ \frac {2 n - 1} {n}} \; \; . 
\end{equation}
Note that $f(x \; ; \; n)$ is a class of functions parametrised
by $n \;$, that $n$ indicates the flatness of $f$ near its
maximum at $x_* \;$, and that $f(x \; ; \; n) \to \frac {2 n}
{x_*} \; (2 x_* - x)$ in the limit $x \to 2 x_* \;$.


\vspace{4ex}

\begin{center}

{\bf 5.  General Features in the LQC -- inspired models}

\end{center}

\vspace{2ex}

Obtaining the cosmological evolution of the universe in the LQC
-- inspired models further requires specifying the equations of
state which give the pressures $p_i$ in terms of $\rho \;$. Once
the equations of state are given, or assumed, equations
(\ref{e1}) -- (\ref{e3}) can be solved for a given set of
initial values $(m^i_0 \;, \; \lambda^i_0)$ at $t = t_0 \;$.
Here and in the following, the $0-$subscripts will denote the
initial values at some initial time $t_0 \;$. Given $m^i_0 \;$,
the values $(f^i_0 \; , \; g_{i 0} \; , \; X_{i 0})$ follow.
Equation (\ref{e1}) then gives $\rho_0 \;$; equations of state
give $p_{i 0} \;$; and, equation (\ref{e2}) gives $(m^i)_{t 0}$
from which the value of $m^i$ at $t_0 \pm \delta t$ follows.
Repeating this procedure will give $m^i \;$ and $(f^i \; , \;
g_i \; , \;X_i)$ for all $t \;$. The initial value $\lambda^i_0$
and equation (\ref{e3}) for $\lambda^i_t$ then determine
$\lambda^i$ for all $t \;$. Thus, equations (\ref{e1}) --
(\ref{e3}) can always be solved numerically.

However, in general, it is not possible to obtain analytical
solutions explicitly. Nevertheless, it is possible to understand
several features of the evolution, as we now describe in a
series of remarks.

\vspace{2ex}

{\bf Remark (1) :} 
Let $f(m^i) = m^i \;$. Then $g_i = 1 \;$, and equation
(\ref{e3}) and the definition of $m^i$ give
\[
\gamma \lambda_{qm} \; \lambda^i_t = m^i 
\; \; , \; \; \;
\hat{c_i} = \gamma L_i \; (a_i)_t \; \; . 
\] 
This shows that $\hat{c}_i$ is related to $(a_i)_t \;$. After a
little algebra, equations (\ref{e1}) and (\ref{e2}) give
equations (\ref{t21}) and (\ref{t22}), the Einstein's equations
for a $(d + 1)$ dimensional homogeneous anisotropic universe.

\vspace{2ex}

{\bf Remark (2) :} 
Equations (\ref{e1}) -- (\ref{e3}) remain invariant under the
shift $m^i \to \tilde{m}^i = m^i + m_s$ where $m_s$ is constant
and same for all $i \;$. Hence, $f(m^i)$ and $f(\tilde{m}^i)$
will lead to the same evolution.

\vspace{2ex}

{\bf Remark (3) :} 
Under the scaling $m^i \to \tilde{m}^i = \alpha \; m^i$ and $t
\to \tilde{t} = \alpha \; t \;$ where $\alpha$ is constant and
same for all $i \;$, and $\rho$ and $p_i$ remain unchanged, we
have
\[
\lambda^i_t \; \to \; \lambda^i_{\tilde{t}} \; = \;
\frac {\lambda^i_t} {\alpha} \; \; , \; \; \; 
\rho_t \; \to \; \rho_{\tilde{t}} \; = \;
\frac {\rho_t} {\alpha} \; \; ,
\]
and, from equations (\ref{xigi}), 
\[
g_i \to \tilde{g}_i = \frac {d f(\tilde{m}^i)} {d \tilde{m}^i} =
\frac {g_i} {\alpha} \; \; , \; \; \;
X_i \to \tilde{X}_i = \frac {X_i} {\alpha} \; \; . 
\]
It then follows that equations (\ref{t23}) and (\ref{e1}) --
(\ref{e3}) remain invariant under these scalings. The
invariances under the shift and the scaling then imply that $\{
f(m^i), \; t \}$ and $\{ f(\alpha m^i + m_s), \; \alpha t \}$
will lead to the same evolution.

Note that these invariance properties are accidental, are not
present even for the more general models presented in
\cite{k16}, and arise from the structure of the equations
(\ref{e1}) -- (\ref{e3}) considered here. Nevertheless, these
properties are useful practically. The scaling with $\alpha = -
1$ is particularly useful for our purposes here. Then $t \to -
t$ and, hence, this scaling may be thought of as reversing the
direction of time and the corresponding evolution may be thought
of as that seen when one goes back in time.

\vspace{2ex}

{\bf Remark (4) :} 
The density $\rho$ and the expansion rates $\lambda^i_t$ remain
finite if the functions $f^i$ and their first derivatives $g_i$
are finite. If all the higher derivatives $\frac {d^n f(m^i)} {d
(m^i)^n}$ are finite then all the higher time derivatives $\frac
{d^n \lambda^i} {d t^n}$ will also be finite. The evolution will
then be non singular.

\vspace{2ex}

{\bf Remark (5) :}
In our models, we require that $f(x) \to x$ as $x \to 0 \;$.
Then, in this limit, the universe will evolve as in Einstein's
theory. Furthermore, in the interval $0 < x < x_r \;$, let
$f$ remain positive and bounded, let all the derivatives of $f$
also remain bounded, and let $f(x) \propto (x_r - x)$ as $x \to
x_r \;$. It then follows from the above Remarks that the
universe will evolve as in Einstein's theory when $m^i \to 0$
for all $i$, and as its time reversed version when $m^i \to x_r$
for all $i$, and the evolution will remain non singular in
between. Note that the properties mentioned above are satisfied
by $f(x) = sin \; x \;$ for which $x_r = \pi \;$, and also by
the class of functions $f(x \; ; \; n)$ given in equation
(\ref{fn}) for which $x_r = 2 x_* \;$. Many such examples may be
constructed easily.

\vspace{2ex}

{\bf Remark (6) :}
Consider a function $f(x)$ which $\to x$ as $x \to 0 \;$. Let
$f$ be positive and bounded, and let all its derivatives also be
bounded, in the interval $0 < x < x_r \;$. Then the universe
will evolve as in Einstein's theory when $m^i \to 0$ for all $i$
and the evolution will remain non singular until $m^i$ approach
$x_r \;$. The nature of the evolution in the limit $m^i \to x_r$
will depend on the behaviour of $f(x)$ as $x \to x_r \;$.

In LQC, the function $f(x) = sin \; x$ is determined by the
underlying theory. In the present LQC -- inspired models, the
generalisation is empirical and no underlying theory is invoked
which may determine $f(x) \;$. This feature is a shortcoming of
the present models. But it may be taken as a strength also. One
may then consider possible asymptotic behaviours of $f(x)$ from
a completely general point of view, obtain a variety of
asymptotics of the evolution, and thereby gain insights into the
various types of singular and non singular evolutions possible.
From such a perspective, it is quite natural to consider the
case where $f(x) \propto (x_r - x)^q $ in the limit $x \to x_r
\;$. Note that it is also possible that $x_r$ is infinite. One
may then also consider the cases where, in the limit $x \to
\infty \;$, the function $f(x) \propto x^q \;$, or $f(x) \to
(const) \;$, or $f(x) \propto e^{- b \; x} \;$.

\vspace{2ex}

{\bf Remark (7) :}
Consider the isotropic case. Let the equation of state be given
by $p = w \; \rho$ where $w$ is a constant and $1 + w > 0 \;$.
Then, equations (\ref{tiso3}) and (\ref{eiso1}) -- (\ref{eiso3})
can be solved explicitly if certain integrations and functional
inversions can be performed. Denoting the initial values with
$0-$subscripts, it follows from equations (\ref{tiso3}) and
(\ref{eiso1}) that
\begin{equation}\label{am}
\frac {\rho} {\rho_0} = \frac {f^2} {f^2_0} = 
\left( \frac {a} {a_0} \right)^{- (1 + w) \; d}
\; \; , \; \; \; \rho_0 = \rho_{qm} \; f^2_0 \; \; , 
\end{equation}
which gives $a$ in terms of $m \;$. Note that if $f$ has a
maximum $f_{mx}$ then the density $\rho$ has a maximum
$\rho_{mx}$ and the scale factor $a$ has a minimum $a_{mn}$
which are given by
\begin{equation}\label{fmx}
\rho_{mx} = \rho_{qm} \; f^2_{mx} \; \; , \; \; \;
a_{mn} = a_0 \; \left( \frac {f_0} {f_{mx}}
\right)^{\frac {2} {(1 + w) \; d}} \; \; . 
\end{equation}
Equations (\ref{eiso1}) and (\ref{eiso2}) give
\begin{equation}\label{calF}
{\cal F} = - \; \int \frac {d m} {f^2} \; = \;
c_{qm} \; \tilde{t}
\; \; \; , \; \; \; \;
\tilde{t} = t - t_0 + \frac {{\cal F}_0} {c_{qm}}
\end{equation}
where $c_{qm} = \frac {(1 + w) \; d} {2 \; \gamma \lambda_{qm}}
\;$. Equation (\ref{calF}) defines ${\cal F}(m)$ and gives $t$
in terms of $m \;$. Inverting $t(m)$ then gives $m(t)$ which, in
turn, gives $a(t) \;$.

\vspace{2ex}

{\bf Remark (8) :} 
Consider the isotropic case with $p = w \; \rho$ as above. Now,
let these $p$ and $\rho$ be due to a minimally coupled scalar
field $\sigma$ with a potential $V(\sigma) \;$, see equation
(\ref{rhopsigma}). Writing $\sigma_t^2 = (1 + w) \rho \;$,
equations (\ref{eiso1}) and (\ref{eiso2}) give
\[
\sigma_t = \left( \sqrt{ (1 + w) \; \rho_{qm} } \right) \; f
\; \; , \; \; \; \;
\frac {d m} {d \sigma} = \frac {m_t} {\sigma_t} = - \; \frac
{\gamma \lambda_{qm} \kappa^2} {d - 1} \; \sigma_t \; \; .
\]
It then follows that 
\begin{equation}\label{calS}
{\cal S} = - \; \int \frac {d m} {f} \; = \;
c_w \; \tilde{\sigma} 
\; \; \; , \; \; \; \;
\tilde{\sigma} = \sigma - \sigma_0 + \frac {{\cal S}_0} {c_w} 
\end{equation}
where $c_w = \sqrt{ \frac {(1 + w) \; \kappa^2 d} {2 \; (d - 1)}
} \;$. Equation (\ref{calS}) defines ${\cal S}(m)$ and gives
$\sigma$ in terms of $m \;$, hence in terms of $t$ if $m(t)$ is
known. Inverting $\sigma(m)$ gives $m(\sigma) \;$, and the
scalar field potential $V(\sigma)$ then follows from
\begin{equation}\label{v2}
2 \; V = (1 - w) \; \rho_{qm} \; f^2 \; \; .
\end{equation}
Note that $V$ given in equation (\ref{v1}) follows by taking
$f(m) = m$ here, or by applying the above manipulations to
equations (\ref{tiso1}) and (\ref{tiso2}).

\vspace{2ex}

{\bf Remark (9) :} 
Consider the isotropic case and consider the evolution near a
maximum of the function $f(m) \;$. Let $m_b$ be a maximum of $f$
and, near its maximum, let
\begin{equation}\label{fmax}
f(m) \; \simeq \; f_{mx} \; \left( 1 - f_1 (m_b - m)^{2 n}
\right)
\end{equation}
where $f_{mx}$ and $f_1$ are positive constants and, as in
equation (\ref{fn}), $n$ is a positive integer which indicates
the flatness of $f$ near its maximum. Let $t_b$ be the time when
$f$ reaches its maximum. Then, as $t \to t_b \;$, it follows
from equations (\ref{calF}), (\ref{am}), and (\ref{calS}) that
\begin{eqnarray}
m_b - m & \simeq & f_{mx}^2 \; c_{qm} \; (t - t_b)
\label{mtb} \\
& & \nonumber \\
a & \simeq & a_{mn} \; \left( 1 + a_1 \; (t - t_b)^{2 n} \right)
\label{atb} \\
& & \nonumber \\
c_w \; (\sigma - \sigma_b) & \simeq &
f_{mx} \; c_{qm} \; (t - t_b) \label{sigmatb} 
\end{eqnarray}
where $\sigma_b$ is a constant, $a_{mn} = a_0 \; \left( \frac
{f_0} {f_{mx}} \right)^{ \frac {2} {(1 + w) d} }$, and $a_1 =
\frac {2 f_1} {(1 + w) d} \; (f_{mx}^2 \; c_{qm})^{2 n} \;$.

\newpage

\vspace{4ex}

\begin{center}

{\bf 6. Isotropic evolutions}

\end{center}

\vspace{2ex}

Consider the isotropic case, with the equation of state given by
$p = w \; \rho$ where $w$ is a constant and $1 + w > 0 \;$. We
will assume that these $p$ and $\rho$ are due to a minimally
coupled scalar field $\sigma$ with a potential $V(\sigma) \;$.
We will first study the examples of functions $f(m)$ for which
equations (\ref{eiso1}) -- (\ref{eiso3}) can be solved
explicitly. Using these examples, we will then study the
asymptotic evolutions for the functions $f(m)$ and the limits
given in Remark {\bf (6)}. As clear from Remarks {\bf (7)} and
{\bf (8)}, obtaining explicit solutions depends on whether the
integrations in equations (\ref{calF}) and (\ref{calS}), and the
consequent functional inversions, can be performed explicitly.


\vspace{4ex}

\begin{center}

{\bf Example I : $\; \; \mathbf f(m) = sin \; m$}

\end{center}

\vspace{2ex}

It turns out that all the required integrations and functional
inversions can be performed for the function $f(m) = sin \; m
\;$. As can be checked easily, ${\cal F}$ and ${\cal S}$ are
given by
\[
{\cal F} = cot \; m \; \; \; , \; \; \; \; \; 
e^{{\cal S}} = cot \; \frac {m} {2} \; \; .
\]
Also, $f$ and ${\cal S}$ may be expressed in terms of ${\cal F} \;$:
\[
f = \frac {1} {\sqrt{ 1 + {\cal F}^2}} \; \; \; , \; \; \; \; \;
Cosh \; {\cal S} = \frac {1} {sin \; m} = \sqrt{ 1 + {\cal F}^2}
\; \; .
\]
Equations (\ref{am}), (\ref{calF}), and (\ref{calS}) now give
the solutions
\begin{equation}\label{asigma}
\left( \frac {a} {a_0} \right)^{(1 + w) \; d} = \frac
{1 + c^2_{qm} \; \tilde{t}^2} {1 + c^2_{qm} \; \tilde{t}^2_0}
\; \; \; , \; \; \; \; \;
Cosh \left( c_w \tilde{\sigma} \right)
= \sqrt{ 1 + c^2_{qm} \; \tilde{t}^2} \; \; . 
\end{equation}
The potential $V(\sigma)$ for the scalar field given in equation
(\ref{v2}) now becomes
\begin{equation}\label{vsinm}
V(\sigma) = \frac {(1 - w) \; \rho_{qm}}
{2 \; Cosh^2 \left( c_w \tilde{\sigma} \right)} \; \; . 
\end{equation}
We note, in passing, that the above expressions are the $(d +
1)$ dimensional generalisation of the $(3 + 1)$ dimensional LQC
solution given in \cite{Vlq}.


\vspace{4ex}

\begin{center}

{\bf Example II :
$\; \; \mathbf f(m) = m^q \; \; \; , \; \; \; m \ge 0$}

\end{center}

\vspace{2ex}

The function $f(m) = m^q$ with $m \ge 0 \;$, by itself, may not
be of much interest since $f$ has no finite maximum and,
generically, the evolution will be singular. However, for this
example, all the integrations in equations (\ref{calF}) and
(\ref{calS}), and the consequent functional inversions, can be
performed which will lead to explicit solutions. Moreover,
together with the shifting and the scaling of $m$ described in
Remarks {\bf (2), (3),} and {\bf (5)}, the solutions for this
example can be used to understand the asymptotics of the
nonsingular evolutions in a variety of cases which may be of
interest. With this purpose in mind, we consider this example
and present the explicit isotropic solutions. It also turns out
that the solutions for the $0 < 2 q < 1$ case exhibit an
interesting feature, and the solutions for the $q = \frac {1}
{2}$ case are quite intriguing.

Before proceeding with the isotropic case, we first note that
when the function $f(x) = x^q \;$, we have $f^i = (m^i)^q \; ,
\; \; m^i g_i = q f^i \;$, and hence
\[
\sum_j m^j X_j = q \; \sum_{j k} G_{j k} f^j f^k \; \; .
\]
Equation (\ref{e2}) then gets simplified and becomes
\begin{equation}\label{e2b} 
(m^i)_t \; + \; \Lambda_t \; m^i \; = \; \gamma \lambda_{qm}
\kappa^2 \; \sum_j G^{i j} \; \left\{ (2 q - 1) \; \rho - p_j
\right\} \; \; .
\end{equation}
If $p_i = p = w \; \rho$ then the above equation becomes
\begin{equation}\label{e2c} 
(m^i)_t \; + \; \Lambda_t \; m^i \; = \; \frac {\gamma
\lambda_{qm} \kappa^2} {d - 1} \; (2 q - 1 - w) \; \rho \; \; .
\end{equation}

Now consider the completely isotropic case given by equation
(\ref{iso}), and the corresponding equations of motion
(\ref{tiso3}) and (\ref{eiso1}) -- (\ref{eiso3}). Note that the
function $g = q \; m^{q - 1}$ now. Hence, equations
(\ref{eiso1}) and (\ref{eiso3}) give
\begin{equation}\label{eiso3q}
H^2 \; = \; \frac {q^2} {\gamma^2 \lambda_{qm}^2} \; \left(
\frac {\rho} {\rho_{qm}} \right)^{\frac {2 q - 1} {q}} \; \; .
\end{equation}
Note also that for any quantity $Z(m)$ which varies as $m^\alpha
\;$, it follows from equations (\ref{eiso1}) and (\ref{eiso2}),
and from $f = m^q \;$, that
\begin{equation}\label{zalpha}
Z_t \; = \; Z_m \; m_t \; \sim \; m^{2 q - 1 + \alpha} 
\; \; \; \longrightarrow \; \; \; 
\frac {d^n Z} {d t^n}  \sim \; m^{n \; (2 q - 1) + \alpha}
\; \; .
\end{equation}
If $Z = (ln \; a)$ then $\alpha = 0$ formally and $\frac {d^n \;
(ln \; a)} {d t^n} \sim \; m^{n \; (2 q - 1)} \;$. In the
following, we will take the evolution to be singular if any of
the time derivatives of $(ln \; a)$ diverges, and to be non
singular otherwise.

The general explicit solution for $f = m^q$ follows upon
performing the integrations in equations (\ref{calF}) and
(\ref{calS}), and the consequent functional inversions. We will
now present these solutions.


\vspace{2ex}

\begin{center}

{$\mathbf q \; \ne \; \frac {1} {2} \;$} 

\end{center}

\vspace{2ex}

For $q \ne \frac {1} {2}$ or $1 \;$, equations (\ref{calF}) and
(\ref{calS}) give
\begin{eqnarray}
m^{1 - 2 q} \; = & (2 q - 1) \; c_{qm} \; (t - t_0) 
+ m_0^{1 - 2 q} &  \equiv \; T \label{mX} \\
& & \nonumber \\
m^{1 - q} \; = & (q - 1) \; c_w \; (\sigma - \sigma_0) 
+ m_0^{1 - q} & \equiv \; A \; \sigma + B \label{mAB}
\end{eqnarray} 
where the constants $A$ and $B$ can be read off easily. It then
follows from equation (\ref{am}) that the scale factor $a$ is
given by
\begin{equation}\label{qam}
\frac {a} {a_0} = \left( \frac {m} {m_0}
\right)^{ - \; \frac {2 q} {(1 + w) d}} = \left( \frac {T} {T_0}
\right)^{ \frac {2 q} {(2 q - 1) \; (1 + w) d}}
\end{equation}
where $T_0 = m_0^{1 - 2 q} \;$. And, it follows from equation
(\ref{v2}) that the scalar field potential $V$ is given by
\begin{equation}\label{vq}
2 \; V(\sigma) = (1 - w) \; \rho_{qm} \; m^{2 q} =
(1 - w) \; \rho_{qm} \; \left( A \; \sigma + B
\right)^{ \frac {2 q} {1 - q} } \; \; .
\end{equation}

For $q = 1 \;$, one obtains the standard Einstein's equations as
described in Remark {\bf (1)}. Thus, for example, one obtains
from equations (\ref{am}) -- (\ref{calS}) that
\[
m \; = \; \frac {1} {c_{qm} \tilde{t}}
\; = \; e^{- c_w \; \tilde{\sigma}} \; \; , \; \; \;
a \; \sim \; m^{- \frac {2} {(1 + w) d}} \; = \; \left(
c_{qm} \tilde{t} \right)^{\frac {2} {(1 + w) d}} \; \; . 
\]
The scalar field potential is given by $V \sim m^2 \sim e^{- 2
\; c_w \; \tilde{\sigma}} \;$, see equation (\ref{v1}).

\vspace{2ex}

The evolution of the universe follows from equations
(\ref{qam}), (\ref{mX}), and (\ref{zalpha}).

\vspace{2ex}

{$\mathbf 2 \; q > 1 :$}
In this case, as $m \to 0 \;$, the scale factor $a \to \infty
\;$, the time $t \to \infty \;$, and the time derivatives of
$(ln \; a)$ will not diverge. The evolution is non singular in
this limit, and proceeds smoothly as $m$ increases further. As
$m \to \infty \;$, the scale factor $a \to 0 \;$ and the time $t
\to t_s$ from above where $t_s$ is finite and its value can be
read off easily but is not important here. Also, the time
derivatives of $(ln \; a)$ diverge and, hence, the evolution is
singular in this limit. Thus, for the $2 q > 1$ case, the
universe starts with a zero size and a singularity at a finite
time in the past, and expands to infinite size in the infinite
future with no further singularities.

\vspace{2ex}

{$\mathbf 0 < 2 q < 1 :$}
In this case, as $m \to 0 \;$, the scale factor $a \to \infty
\;$ and the time $t \to t_s$ from below where $t_s$ is finite
and its value can be read off easily but not important here.
Also, the time derivatives of $(ln \; a)$ diverge and, hence,
the evolution is singular in this limit. The evolution proceeds
smoothly as $m$ increases further. As $m \to \infty \;$, the
scale factor $a \to 0 \;$, the time $t \to - \infty \;$, and the
time derivatives of $(ln \; a)$ will not diverge. The evolution
is non singular in this limit. Thus, for the $0 < 2 q < 1$ case,
universe approaches a zero size in the infinite past but with no
singularity, and expands to infinite size with a singularity at
a finite time in the future.

Note the interesting feature that there is a singularity when
the universe is increasing to infinite size, and no singularity
when it is contracting to zero size. Such a feature is opposite
to what one usually comes across in Einstein's theory.

\vspace{2ex}

{$\mathbf q < 0 :$}
In this case, as $m \to 0 \;$, the scale factor $a \to 0 \;$ and
the time $t \to t_s$ from below where $t_s$ is finite and its
value can be read off easily but not important here. Also, the
time derivatives of $(ln \; a)$ diverge and, hence, the
evolution is singular in this limit. The evolution proceeds
smoothly as $m$ increases further. As $m \to \infty \;$, the
scale factor $a \to \infty \;$, the time $t \to - \infty \;$,
and the time derivatives of $(ln \; a)$ will not diverge. The
evolution is non singular in this limit. Thus, for the $q < 0$
case, the universe is infinite in size in the infinite past, and
contracts to zero size and a singularity at a finite time in the
future.


\vspace{2ex}

\begin{center}

{$\mathbf q \; =  \; \frac {1} {2} $} 

\end{center}

\vspace{2ex}

Consider now the $q = \frac {1} {2} \;$ case. Equations
(\ref{calF}) and (\ref{am}) give
\begin{eqnarray}
m & = & m_0 \; e^{ - \; c_{qm} \; (t - t_0)}  \label{mthalf} \\
& & \nonumber \\
\frac {a} {a_0} & = & \left( \frac {m} {m_0}
\right)^{ - \; \frac {2 q} {(1 + w) d}} \; = \; \;
e^{ \frac {t - t_0} {2 \gamma \lambda_{qm}} }
\label{athalf}
\end{eqnarray}
where we have used $c_{qm} = \frac {(1 + w) \; d} {2 \; \gamma
\lambda_{qm}} \;$ in the last equality. Equations (\ref{mAB})
and (\ref{vq}) remain valid, and note that equation (\ref{vq})
gives $V \propto (A \sigma + B)^2 \;$.

In this case, as $m \to 0 \;$, the scale factor $a \to \infty
\;$ and the time $t \to \infty \;$. As $m \to \infty \;$, the
scale factor $a \to 0 \;$ and the time $t \to - \infty \;$.
Also, clearly, the time derivatives of $(ln \; a)$ will not
diverge and, hence, the evolution is non singular. Thus, for the
$q = \frac {1} {2} $ case, the universe starts with a zero size
in the infinite past, and expands to infinite size in the
infinite future with no singularities.

Note that the evolution in the $q = \frac {1} {2} \;$ case
straddles the border between the singular and non singular
evolution : in the limit $m \to 0 \;$, the evolution is non
singular for $2 q > 1$ and is singular for $2 q < 1 \;$; in the
limit $m \to \infty \;$, the evolution is singular for $2 q > 1$
and is non singular for $2 q < 1 \;$; for the border case $2 q =
1 \;$, the evolution is non singular in both the asymptotic
limits.

Also note that, for $q = \frac {1} {2} \;$, we have an
exponentially growing scale factor with the exponential rate set
by the parameter $\lambda_{qm}$ alone, which is related to the
quantum of the $(d - 1)$ dimensional area as in LQC. The density
is given by equation (\ref{am}),
\[
\rho \; \sim \; f^2 \; \sim \; a^{- (1 + w) \; d}
\; \sim \; e^{- c_{qm} t} \; \; ,
\]
and does not remain constant. Also, there is no restriction on
$w$, the equation of state parameter. So, this exponential
growth of the scale factor is unlike that which occurs in
Einstein's theory due to a positive cosmological constant. See
equation (\ref{eiso3q}), now with $2 q = 1 \;$, to see how this
comes about. These results are intriguing and fascinating but
their significance, if any, is not clear to us at present.


\vspace{4ex}

\begin{center}

{\bf Example III : $\; \; \mathbf f(m) = e^m $}

\end{center}

\vspace{2ex}

We consider the function $f(m) = e^m \;$ for the same reasons as
given for Example II : solutions can be obtained explicitly.
They may then be used to understand the asymptotics of the
nonsingular evolutions in other cases which may be of
interest. For this example, equations (\ref{eiso1}) and
(\ref{eiso3}) give
\begin{equation}\label{eiso3ex}
H^2 \; = \; \frac {1} {\gamma^2 \lambda_{qm}^2} \; \left(
\frac {\rho} {\rho_{qm}} \right)^2 \; \; . 
\end{equation}
Equations (\ref{calF}) and (\ref{calS}) give
\begin{eqnarray}
e^{- 2 m} \; = & 2 \; c_{qm} \; (t - t_0) + e^{- 2 m_0} &
\equiv \; T \label{mXex} \\
& & \nonumber \\
e^{- m} \; = & \; c_w \; (\sigma - \sigma_0) + e^{- m_0} &
\equiv \; A \; \sigma + B  \label{mABex}
\end{eqnarray} 
where the constants $A$ and $B$ can be read off easily. It then
follows from equation (\ref{am}) that the scale factor $a$ is
given by
\begin{equation}\label{amex}
\frac {a} {a_0} = e^{ - \; \frac {2 (m - m_0)} {(1 + w) d}}
= \left( \frac {T} {T_0} \right)^{ \frac {1} {(1 + w) d}}
\end{equation}
where $T_0 = e^{- 2 m_0} \;$. And, it follows from equation
(\ref{v2}) that the scalar field potential $V$ is given by
\begin{equation}\label{vex}
2 \; V(\sigma) \; = \; (1 - w) \; \rho_{qm} \; e^{2 m} \; = \;
\frac {(1 - w) \; \rho_{qm}} {\left( A \; \sigma + B
\right)^2} \; \; .
\end{equation}
Also, for any quantity $Z(m)$ which varies as $m^\alpha \;$, it
follows from equations (\ref{eiso1}) and (\ref{eiso2}), and from
$f = e^m \;$, that
\begin{equation}\label{zex}
\frac {d^n Z} {d t^n} \sim \; \left( \sum_{k = 1}^n A_k \;
m^{\alpha - k} \right) \; e^{2 n \; m} 
\end{equation}
where $A_k$ are some constants. The evolution of the universe
follows from equations (\ref{amex}), (\ref{mXex}), and
(\ref{zex}).

\vspace{2ex}

As $m \to - \infty \;$, the scale factor $a \to \infty \;$, the
time $t \to \infty \;$, and the time derivatives of $(ln \; a)$
will not diverge. The evolution is non singular in this limit,
and proceeds smoothly as $m$ increases further. As $m \to \infty
\;$, the scale factor $a \to 0 \;$ and the time $t \to t_s$ from
above where $t_s$ is finite and its value can be read off easily
but not important here. Also, the time derivatives of $(ln \;
a)$ diverge and, hence, the evolution is singular in this
limit. Thus, the universe starts with a zero size and a
singularity at a finite time in the past, and expands to
infinite size in the infinite future with no further
singularities.

In {\bf Table I} below, we tabulate the asymptotic behaviour of
the scale factor $a(t) \;$ in Examples I -- III, described above
in detail. In the Table, we give the forms of the function
$f(m)$ and the asymptotic values of $a(t)$ in the limit $t \to
\pm \infty \;$ or $t_s \;$ where $t_s$ is finite. Also, we use
the letter $S$ or $NS$ to denote whether the evolution in this
limit is singular or non singular.


\vspace{4ex}


\vspace{2ex}

\begin{tabular}{||c||c|c||c|c|c||} 
\hline \hline 
& & & & & \\ 

&
$f(m)$ &
& 
$t \; \to \; - \; \infty$ & 
$t \; \to \; t_s$ & 
$t \; \to \; \infty$ \\ 
& & & & & \\ 
\hline  \hline 

& & & & & \\ 

I &
$sin \; m$ & 
& 
$\infty \; , \; NS $ &
$\to \; a_{min} \; \to $ & 
$\infty \; , \; NS $ \\

& & & & & \\ 
\hline  \hline 

& & & & & \\ 

&
& 
$2 q \; > \; 1$ &
&
$0 \; , \; S $ & 
$\infty \; , \; NS $ \\

& & & & & \\ 
\cline{3-6}  
& & & & & \\ 

& 
& 
$2 q \; = \; 1$ &
$0 \; , \; NS $ & 
$\longrightarrow$ &
$\infty \; , \; NS $ \\

II & $m^q$ & & & & \\ 
\cline{3-6}  
& & & & & \\ 

&
& 
$0 \; < \; 2 q \; < \; 1$ &
$0 \; , \; NS $ & 
$\infty \; , \; S $ & 
\\

& & & & & \\ 
\cline{3-6}  
& & & & & \\ 

&
&
$q \; < \; 0$ &
$\infty \; , \; NS $ & 
$0 \; , \; S $ & 
\\

& & & & & \\ 
\hline  \hline 

& & & & & \\ 
III &
$e^m$ & 
& 
& 
$0 \; , \; S $ & 
$\infty \; , \; NS $ \\

& & & & & \\ 
\hline  \hline 

\end{tabular}

\vspace{2ex}

\begin{center}

{\bf Table I : Asymptotic behaviours of the scale factor
$\mathbf a(t) \;$.}

{\em Tabulated here are the forms of the function $f(m)$ in
Examples I -- III, and the asymptotic values of the scale factor
$a(t)$ in the limit $t \to \pm \infty \;$ or $t_s \;$ where
$t_s$ is finite. The accompanying letter $S$ or $NS$ denotes
whether the evolution in this limit is singular or non
singular. The entire evolution is non singular only in Example
I, where it has a bounce, and in Example II with $ 2 q = 1$,
where it has no bounce. The values of $m$ in these limits follow
straightforwardly and, hence, are not tabulated here.}

\end{center} 


\vspace{4ex}

\begin{center}

{\bf Other Examples} 

\end{center}

\vspace{2ex} 

Using Examples II and III, we now study the asymptotic
evolutions for the functions $f(m)$ and the limits given in
Remark {\bf (6)}. For these functions, we assume that $f(m) \to
m$ in the limit $m \to 0 \;$ and also that, until $m$ approaches
the limit of interest, $f$ remains positive and bounded, and all
its derivatives also remain bounded. It then follows that, in
the limit $m \to 0 \;$, the evolution is as in Einstein's
theory, the scale factor $a \to \infty \;$, the time $t \to
\infty \;$, and the time derivatives of $(ln \; a)$ will not
diverge. As $m$ increases from $0 \;$, the time $t$ decreases
from $\infty \;$, the scale factor $a$ decreases from $\infty
\;$ but remains non zero since $f$ remains bounded from above,
and the evolution proceeds smoothly with no singularity until
$m$ approaches the limit of interest.

We characterise the evolution in these limits of interest as
follows. If the time derivatives of $(ln \; a)$ do not diverge
then the evolution will be referred to as non singular; if the
scale factor $a \to \infty$ then the evolution will be said to
have a bounce; when there is a bounce, if the evolution as $a
\to \infty$ is same as that in Einstein's theory then the
evolution will be referred to as symmetric; and, when there is a
bounce, if the evolution as $a \to \infty$ is different from
that in Einstein's theory then evolution will be referred to as
asymmetric.

Note that the words symmetric and asymmetric refer not to the
actual shape of $a(t) \;$, but refer only to whether or not
$a(t)$ evolves as in Einstein's theory at both the ends of a
bounce where $a \to \infty \;$. Thus, for example, if $f(m) =
sin \; m \;$, or if $f(m) = f(m \; ; \; n) \;$ given in equation
(\ref{fn}), then the evolution is non singular, has a bounce,
and is symmetric. It turns out that, because of the symmetric
shapes of $sin \; m \;$ and $f(m \; ; \; n) \;$, the shape of
$a(t)$ is also symmetric but, generically, this need not be the
case.

\newpage

\vspace{4ex}

\begin{center}

{\bf Example IV : $\; \; \mathbf f(m) \; \propto \; (m_s - m)^q
\; \; \; as \; \; \; m \; \to \; m_s$}

\end{center}

\vspace{2ex}

We now consider the Example where the function $f(m) \; \propto
\; (m_s - m)^q \;$ in the limit of interest $m \to m_s \;$. The
evolution in the limit $m \to m_s$ can be read off from the
asymptotic behaviour in Example II as $m \to 0$ there. One also
needs to change the sign of $t \;$ there since $f \propto (m_s -
m)^q$ now. It then follows that if $2 q \ge 1 \;$ then, as $m
\to m_s \;$, the scale factor $a \to \infty \;$, the time $t \to
- \infty \;$, and the time derivatives of $(ln \; a)$ will not
diverge. Hence, the evolution is non singular and is different
from that in Einstein's theory unless $q = 1 \;$.  Thus, as $m$
increases from $0$ to $m_s \;$, the universe evolves as in
Einstein's theory in the limit $t \to \infty \;$, the scale
factor $a$ remains non zero throughout, increases to $\infty$
and, unless $q = 1 \;$, evolves asymmetrically in the limit $t
\to - \infty \;$. The evolution has a bounce, is asymmetric
unless $q = 1 \;$, and remains non singular throughout.

If $0 < 2 q < 1 \;$ then, as $m \to m_s \;$, the scale factor $a
\to \infty \;$, the time $t \to t_s$ which is finite, the time
derivatives of $(ln \; a)$ diverge and, hence, the evolution is
singular. Thus, as $m$ increases from $0$ to $m_s \;$, the
universe evolves as in Einstein's theory in the limit $t \to
\infty \;$, the scale factor $a$ remains non zero throughout and
increases to $\infty$ at a finite time $t_s$ in the past. The
evolution has a bounce, and is singular as $t \to t_s \;$.

If $q < 0 \;$ then, as $m \to m_s \;$, the scale factor $a \to 0
\;$, the time $t \to t_s$ which is finite, the time derivatives
of $(ln \; a)$ diverge and, hence, the evolution is singular.
Thus, as $m$ increases from $0$ to $m_s \;$, the universe
evolves as in Einstein's theory in the limit $t \to \infty \;$,
the scale factor $a$ decreases from $\infty$ to $0$ at a finite
time $t_s$ in the past, and the evolution is singular as $t \to
t_s \;$.


\vspace{4ex}

\begin{center}

{\bf Example V : $\; \; \mathbf f(m) \; \propto \; m^q
\; \; \; as \; \; \; m \; \to \; \infty$}

\end{center}

\vspace{2ex}

We now consider the Example where the function $f(m) \propto m^q
\;$ in the limit of interest $m \to \infty \;$. The evolution in
the limit $m \to \infty$ can be read off from the asymptotic
behaviour in Example II as $m \to \infty$ there. It then follows
that if $2 q > 1 \;$ then, as $m \to \infty \;$, the scale
factor $a \to 0 \;$, the time $t \to t_s$ which is finite, the
time derivatives of $(ln \; a)$ diverge and, hence, the
evolution is singular. Thus, as $m$ increases from $0$ to
$\infty \;$, the universe evolves as in Einstein's theory in the
limit $t \to \infty \;$, the scale factor $a$ decreases from
$\infty$ to $0$ at a finite time $t_s$ in the past, and the
evolution is singular as $t \to t_s \;$.

If $0 < 2 q \le 1 \;$ then, as $m \to \infty \;$, the scale
factor $a \to 0 \;$, the time $t \to - \infty \;$, the time
derivatives of $(ln \; a)$ will not diverge and, hence, the
evolution is non singular. Thus, as $m$ increases from $0$ to
$\infty \;$, the universe evolves as in Einstein's theory in the
limit $t \to \infty \;$, the scale factor $a$ decreases from
$\infty$ to $0$ as $t \to - \infty \;$, and the evolution
remains non singular throughout.

If $q < 0 \;$ then, as $m \to \infty \;$, the scale factor $a
\to \infty \;$, the time $t \to - \infty \;$, the time
derivatives of $(ln \; a)$ will not diverge and, hence, the
evolution is non singular. Thus, as $m$ increases from $0$ to
$\infty \;$, the universe evolves as in Einstein's theory in the
limit $t \to \infty \;$, the scale factor $a$ decreases from
$\infty$ to some non zero value and then increases again to
$\infty$ as $t \to - \infty \;$, the evolution has a bounce, and
remains non singular throughout.


\vspace{4ex}

\begin{center}

{\bf Example VI : $\; \; \mathbf f(m) \; \propto \; 1 \;
\; \; as \; \; \; m \; \to \; \infty \;$}

\end{center}

\vspace{2ex}

We now consider the Example where the function $f(m) \propto 1
\;$ in the limit of interest $m \to \infty \;$. This Example can
be thought of as a special case of Example V with $q = 0 \;$. It
follows straightforwardly from equations (\ref{am}) and
(\ref{calF}) that the scale factor $a \to (const) \;$ and the
time $t \to - \infty \;$. Clearly, the time derivatives of $(ln
\; a)$ will not diverge and, hence, the evolution is non
singular. Thus, as $m$ increases from $0$ to $\infty \;$, the
universe evolves as in Einstein's theory in the limit $t \to
\infty \;$, the scale factor $a$ decreases from $\infty$ to some
non zero constant value as $t \to - \infty \;$, and the
evolution remains non singular throughout.

The density $\rho$ also approaches a non zero constant value as
$t \to - \infty \;$. This phase of the evolution is then similar
to what is expected in string/M theory where, as one goes back
in time, the ten/eleven dimensional early universe is believed
to enter and remain in a Hagedorn phase in which its temperature
is of the order of $l_s^{- 1} \;$ and its density is of the
order of $l_s^{- (d + 1)} \;$ where $l_s$ is the string length
scale \cite{bowick} -- \cite{k06}. 

\newpage

\vspace{4ex}

\begin{center}

{\bf Example VII : $\; \; \mathbf f(m) \; \propto \; e^{- b m}
\; \; \; as \; \; \; m \; \to \; \infty \; \; ; \; \; \; b \; >
\; 0$}

\end{center}

\vspace{2ex}

We now consider the Example where the function $f(m) \propto
e^{- b m} \;$ with $b > 0 \;$, in the limit of interest $m \to
\infty \;$. The evolution in the limit $m \to \infty$ here can
be read off from the asymptotic behaviour in Example III as $m
\to - \infty$ there. One also needs to change the sign of $t \;$
there since $f \propto e^{- b m}$ now. It then follows that, as
$m \to \infty \;$, the scale factor $a \to \infty \;$, the time
$t \to - \infty \;$, the time derivatives of $(ln \; a)$ will
not diverge and, hence, the evolution is non singular.  Thus, as
$m$ increases from $0$ to $\infty \;$, the universe evolves as
in Einstein's theory in the limit $t \to \infty \;$, the scale
factor $a$ decreases from $\infty$ to some non zero value and
then increases again to $\infty$ as $t \to - \infty \;$, the
evolution has a bounce, and remains non singular throughout.


\vspace{4ex}

\begin{center}

{\bf Summary of Examples IV -- VII}

\end{center}

\vspace{2ex}

In Table II below, we tabulate the asymptotic behaviour of the
scale factor $a(t) \;$ in Examples IV -- VII, described above in
detail. In the Table, we give the forms of the function $f(m)$
in the limits of interest and the asymptotic values of $a(t)$ in
the limit $t \to \pm \infty \;$ or $t_s \;$ where $t_s$ is
finite. Also, we use the letter $S$ or $NS$ to denote whether
the evolution in this limit is singular or non singular.

We now highlight the results of Examples IV -- VII by
specifically pointing out the cases of non singular evolutions.
In all these Examples, by assumption, we have that as $m$
increases from $0 \;$, the universe evolves as in Einstein's
theory in the limit $t \to \infty \;$, the scale factor $a$
decreases from $\infty \;$ and remains non zero until $m \to m_s
\;$, or $m \to \infty \;$ as the case may be.

\begin{itemize}

\item

In the Example where $f(m) \propto (m_s - m)^q \;$ as $m \to m_s
\;$, the evolution is non singular and has a bounce if $ 2 q \ge
1 \;$. It is asymmetric unless $q = 1 \;$. Also, $t \to - \infty
\;$ and $a \to \infty \;$ in the limit $m \to m_s \;$.

\item

In the Example where $f(m) \propto m^q \;$ as $m \to \infty \;$,
the evolution is non singular and asymmetric if $ 2 q \le 1 \;$.
Also, $t \to - \infty \;$; and $a \to 0 \;$ if $0 < 2 q \le 1
\;$, $\; a \to (const) \;$ if $q = 0 \;$, and $a \to \infty \;$
if $q < 0 \;$ in the limit $m \to \infty \;$. The evolution for
the $q = 0$ case in this limit is similar to that expected in
the Hagedorn phase of string/M theory.

\item

In the Example where $f(m) \propto e^{- b m} \;$ with $b > 0 \;$
as $m \to \infty \;$, the evolution is non singular, has a
bounce, and is asymmetric. Also, $t \to - \infty \;$ and $a \to
\infty \;$ in the limit $m \to \infty \;$.

\end{itemize} 


\vspace{4ex}


\vspace{2ex}

\begin{tabular}{||c||c|c||c|c|c||} 
\hline \hline 
& & & & & \\ 

&
$f(m)$ &
& 
$t \; \to \; - \; \infty \;$, & 
$t \; \to \; t_s$ & 
$t \; \to \; \infty$ \\ 
& & & & & $all \; NS$ \\  
\hline  \hline 

& & & & & \\

IV &
$(m_s - m)^q \; $, & 
$2 q \; \ge \; 1$ & 
$\infty \; , \; NS $ &
$\to \; a_{min} \; \to $ & 
$\infty$ \\

& & & & & \\ 
\cline{3-6}  
& & & & & \\ 

&
$m \; \to \; m_s \;$, & 
$0 \; < \; 2 q \; < \; 1$ &
&
$\infty \; , \; S $ & 
$\infty$ \\

& & & & & \\ 
\cline{3-6}  
& & & & & \\ 

&
$m_s \; : \; finite$ &
$q \; < \; 0 $ &
&
$0 \; , \; S $ & 
$\infty$ \\

& & & & & \\ 
\hline  \hline 

& & & & & \\ 

V &
$m^q \; $, & 
$2 q \; > \; 1$ & 
&
$0 \; , \; S $ &
$\infty$ \\

& & & & & \\ 
\cline{3-6}  
& & & & & \\ 

&
$m \; \to \; \infty$ & 
$0 \; < \; 2 q \; \le \; 1$ &
$0 \; , \; NS$ &
$\longrightarrow$ &
$\infty$ \\

& & & & & \\ 
\cline{3-6}  
& & & & & \\ 

&
& 
$q \; < \; 0 $ &
$\infty \; , \; NS $ & 
$\to \; a_{min} \; \to$ & 
$\infty$ \\

& & & & & \\ 
\cline{1-1}  \cline{3-6}  
& & & & & \\ 

VI &
& 
$q \; = \; 0 $ &
$const \; , \; NS $ & 
$\longrightarrow $ & 
$\infty$ \\

& & & & & \\ 
\hline  \hline 

& & & & & \\ 
VII &
$e^{- b m}$ & 
$b > 0 \;$ & 
$\infty \; , \; NS$ & 
$\to \; a_{min} \; \to$ & 
$\infty$ \\

&
$m \; \to \; \infty$ & 
& & & \\ 
\hline  \hline 

\end{tabular}

\vspace{2ex}

\begin{center}

{\bf Table II : Asymptotic behaviours of the scale factor
$\mathbf a(t) \;$.}

{\em Tabulated here are the forms of the function $f(m)$ in the
limits of interest in Examples IV -- VII, and the asymptotic
values of the scale factor $a(t)$ in the limit $t \to \pm \infty
\;$ or $t_s \;$ where $t_s$ is finite. In the limit $m \to 0 \;$
in these Examples, $f(m) \to m \;$, $\; t \to \infty \;$, $\;
a(t) \to \infty \;$ as in Einstein's theory, and the evolution
is non singular. The accompanying letter $S$ or $NS$ denotes
whether the evolution in this limit is singular or non singular.
The entire evolution is non singular in several cases in these
Examples.}

\end{center}


\vspace{4ex}

\begin{center}

{\bf 7. Anisotropic evolutions}

\end{center}

\vspace{2ex}

Consider the anisotropic evolutions. The scale factors $a_i =
e^{\lambda^i} \;$ are now different for different $i \;$. Let
the equations of state be given by $p_i = p = w \; \rho \;$
where $w$ is a constant and $1 + w > 0 \;$. The evolution is
described by equations (\ref{tiso3}) and (\ref{e1}) --
(\ref{e4b}). In general, we can not solve these equations
explicitly for any non trivial function $f(x) \;$. Therefore, we
consider Example II where $f(x) = x^q \;$ and study mainly the
asymptotic evolutions in the limit $m^i \to 0 \;$ for all $i
\;$, and in the limit $m^i \to \infty \;$ for all $i \;$. Other
Examples and other limits may be studied similarly.

When $f(x) = x^q$ and $p_i = p = w \; \rho \;$, we have $f^i =
f(m^i) = (m^i)^q \;$ and 
\begin{equation}\label{wrho}
\rho = \rho_0 \; e^{- \; (1 + w) \; (\Lambda - \Lambda_0)}
\; \; , \; \; \;
X_j = q \; (m^j)^{q - 1} \; \sum_k G_{j k} (m^k)^q 
\end{equation}
which follow from equations (\ref{tiso3}) and (\ref{xigi}).
Equations (\ref{e1}), (\ref{e2}) or (\ref{e2c}), and (\ref{e3})
become
\begin{eqnarray}
\sum_{i j} G_{i j} (m^i)^q (m^j)^q & = & 2 \; \gamma^2
\lambda_{qm}^2 \kappa^2 \; \rho_0 \;
e^{- \; (1 + w) \; (\Lambda - \Lambda_0)} \label{qe1} \\
& & \nonumber \\
(m^i)_t \; + \; \Lambda_t \; m^i & = & \frac {\gamma
\lambda_{qm} \kappa^2} {d - 1} \; (2 q - 1 - w) \; \rho_0 \;
e^{- \; (1 + w) \; (\Lambda - \Lambda_0)} \label{qe2} \\
& & \nonumber \\
(\gamma \lambda_{qm}) \; \lambda^i_t & = & q \; 
\sum_{j k} G^{i j} G_{j k} \; (m^j)^{q - 1} \; (m^k)^q 
\; \; . \label{qe3} \\
& & \nonumber 
\end{eqnarray}
One also has $m^i - m^j = \mu^{i j} \; e^{- \Lambda} \;$ where
$\mu^{i j}$ are constants, see equation (\ref{e4b}).


\vspace{2ex} 

\centerline{$\mathbf w \; = \; 2 q - 1 \;$}

\vspace{2ex} 

Explicit solutions can be obtained for the $w = 2 q - 1 \;$
case. In this case, equation (\ref{qe2}) gives
\begin{equation}\label{w=e2}
m^i \; = \; m^i_0 \; e^{- (\Lambda - \Lambda_0)}
\end{equation}
where $m^i_0$ are constants. Consider equation (\ref{qe1}). When
$2 q = 1 + w \;$, the $e^\Lambda -$dependent factors cancel in
this equation and one obtains
\begin{equation}\label{w=e1}
\sum_{i j} G_{i j} (m^i_0)^q (m^j_0)^q \; = \; 2 \; \gamma^2
\lambda_{qm}^2 \kappa^2 \; \rho_0 \; \; .
\end{equation}
Consider equation (\ref{qe3}). Using equation (\ref{w=e2}), one
obtains 
\begin{equation}\label{w=e3}
\lambda^i_t \; = \; \lambda^i_{t0} \; e^{- \; (2 q - 1) \; 
(\Lambda - \Lambda_0)}
\; \; , \; \; \; 
\Lambda_t \; = \; \Lambda_{t0}  \; e^{- \; (2 q - 1) \;
(\Lambda - \Lambda_0)}
\end{equation}
where $\lambda^i_{t0}$ and $\Lambda_{t0}$ are given by 
\begin{equation}\label{kik}
(\gamma \lambda_{qm}) \; \lambda^i_{t0} \; = \; q \; \sum_{j k}
G^{i j} G_{j k} \; (m^j_0)^{q - 1} \; (m^k_0)^q \; \; , \; \; \;
\Lambda_{t0} = \sum_i \lambda^i_{t0} \; \; .
\end{equation}
Solving the equation for $\Lambda_t \;$ in (\ref{w=e3}), one
obtains
\begin{equation}\label{l}
e^{(2 q - 1) \; (\Lambda - \Lambda_0)} \; = \;
(2 q - 1) \; \Lambda_{t0} \; \tilde{t} \; \; , \; \; \;
\tilde{t} = t - t_0 + \frac {1} {(2 q - 1) \; \Lambda_{t0}}
\; \; .
\end{equation}
Then it follows that $\tilde{t}_0 = \frac {1} {(2 q - 1)
\Lambda_{t0}} \;$, 
\begin{equation}\label{lil}
\lambda^i_t \; = \; \frac {\alpha^i} {\tilde{t}} 
\; \; , \; \; \;
e^{\lambda^i - \lambda^i_0} \; = \; \left(
\frac {\tilde{t}} {\tilde{t}_0} \right)^{\alpha^i}
\; \; , \; \; \; \alpha^i \; = \;
\frac {\lambda^i_{t0}} {(2 q - 1) \; \Lambda_{t0}} \; \; ,
\end{equation}
and that
\begin{equation}\label{alphaialpha}
e^{\Lambda - \Lambda_0} \; = \; \left(
\frac {\tilde{t}} {\tilde{t}_0} \right)^\alpha \; \; , \; \; \;
\alpha \; = \;  \sum_i \alpha^i \; = \; \frac {1} {2 q - 1} 
\; \; .
\end{equation}
These are the explicit anisotropic solutions for the $w = 2 q -
1 \;$ case. They are parametrised by the initial values $m^i_0$
and $\lambda^i_0 \;$, which then determine the remaining initial
values $\rho_0 \; , \; \Lambda_0 \; , \; \lambda^i_{t0} \;$, and
$\Lambda_{t0} \;$. The following features of these solutions can
now be seen easily.

\begin{itemize}

\item 

Setting $m^i_0 = m_0$ for all $i$ in the above expressions gives
$\lambda^i_{t0} = \frac {\Lambda_{t0}} {d} \;$ and $\alpha^i =
\frac {1} {(2 q - 1) \; d} \;$, and leads to the solutions given
in Example II with $2 q = 1 + w \;$ now; compare equations
(\ref{qam}) and (\ref{lil}).

\item

When $2 q = 1 \;$, hence $w = 0 \;$, equation (\ref{w=e3}) leads
to
\[
e^{\lambda^i - \lambda^i_0} \; = \;
e^{\lambda^i_{t0} \; (t - t_0)} \; \; , \; \; \;
e^{\Lambda - \Lambda_0} \; = \; e^{\Lambda_{t0} \; (t - t_0)}
\]
where $\lambda^i_{t0}$ and $\Lambda_{t0}$ are given by equation
(\ref{kik}). Also, if $m^i_0 = m_0$ for all $i$ then
$\lambda^i_{t0} = \frac {q} {\gamma \lambda_{qm}} \;$. The
solutions given in equation (\ref{athalf}) then follow. 

\item

Vacuum solutions follow upon setting $\rho_0 = 0 \;$ in equation
(\ref{w=e1}).

\item

The above anisotropic solutions for the $w = 2 q - 1 \;$ case
are the analogs of the standard Kasner--type solutions in
Einstein's theory. Indeed, for $q = 1 \;$, we have $w = 1$ and
equations (\ref{kik}), (\ref{alphaialpha}), (\ref{w=e1}), and
(\ref{lil}) give
\[
\gamma \lambda_{qm} \; \lambda^i_{t0} \; = \; m^i_0
\; \; , \; \; \; \sum_i \alpha^i \; = \; 1 \; \; , \; \; \;
1 - \sum_i (\alpha^i)^2 \; = \; \frac {2 \kappa^2 \; \rho_0}
{\Lambda_{t0}^2} \; \; .
\]
For the vacuum case, $\rho_0 = 0$ and one obtains $\sum_i
\alpha^i = \sum_i (\alpha^i)^2 = 1 \;$.

\end{itemize}


\vspace{2ex} 

\centerline{$\mathbf w \; \ne \;  2 q - 1 \;$}

\vspace{2ex} 

When $w \ne 2 q - 1 \;$, we can not solve equations (\ref{qe1})
-- (\ref{qe3}) explicitly. Therefore, for these cases, we study
only the asymptotic evolutions in the limit $m^i \to 0 \;$ for
all $i \;$, and in the limit $m^i \to \infty \;$ for all $i \;$.
In the asymptotic limits, depending on the value of $q$ and
depending on whether $m^i \to 0$ or $m^i \to \infty \;$ for all
$i \;$, the time $t \to \pm \; \infty \;$ or, after
incorporating a finite shift, $t \to 0 \;$. Below, for the sake
of simplicity, we further restrict our study to only those cases
where $2 q \ne 1 \;$ and where $t \to \infty \;$ or $t \to 0
\;$. Other cases can be studied similarly. In these asymptotic
limits, let $m^i \;$, $e^{\lambda^i} \;$, and $\lambda^i_t$ be
given by the ansatz
\begin{equation}\label{mili}
m^i \; = \; \frac {c^i} {t^b} \; \; , \; \; \; \;
e^{\lambda^i} \; \propto \; t^{\alpha^i}
\; \; \; \longleftrightarrow \; \; \; \;
\lambda^i_t \; = \; \frac {\alpha^i} {t} 
\end{equation}
where $b \; , \; c^i \;$, and $\alpha^i \;$ are constants which
must be determined consistently by equations (\ref{qe1}) --
(\ref{qe3}). The present ansatz gives
\begin{equation}\label{fixi}
f^i \; = \; \frac {(c^i)^q} {t^{b \; q}} \; \; , \; \; \; \;
X_j \; = \; \frac {x_j} {t^{b \; (2 q - 1)}} \; \; , \; \; \; \;
\Lambda_t \; = \; \frac {\alpha} {t} 
\end{equation}
where 
\begin{equation}\label{xialpha}
x_j \; = \; q \; (c^j)^{q - 1} \; \sum_k G_{j k} (c^k)^q
\; \; , \; \; \; \alpha \; = \; \sum_i \alpha^i \; \; .
\end{equation}
We then have, since $\rho \propto e^{- (1 + w) \Lambda} \;$,
\begin{equation}\label{rhoT}
e^\Lambda \; = \; c_\Lambda \; t^\alpha
\; \; , \; \; \; \;
\rho \; = \; \frac {c_\rho} {t^{(1 + w) \; \alpha}} 
\end{equation}
where $c_\Lambda$ and $c_\rho$ are constants. Equations
(\ref{e4b}) and (\ref{qe1}) -- (\ref{qe3}) now give 
\begin{eqnarray} 
\frac {c^i - c^j} {t^b} & = & \frac {\mu^{i j}}
{c_\Lambda \; t^\alpha}  \label{1e4b} \\
& & \nonumber \\
\frac {\sum_{i j} G_{i j} (c^i)^q (c^j)^q} {t^{2 b q}} & = &
\frac {2 \; \gamma^2 \lambda_{qm}^2 \kappa^2 \; c_\rho}
{t^{(1 + w) \; \alpha}} \label{1e1} \\
& & \nonumber \\ 
\frac {(\alpha - b) \; c^i} {t^{1 + b}} & = & 
\frac {\gamma \lambda_{qm} \kappa^2 \; (2 q - 1 - w) \; c_\rho}
{(d - 1) \; t^{(1 + w) \; \alpha}}  \label{1e2c} \\
& & \nonumber \\ 
(\gamma \lambda_{qm}) \; \frac {\alpha^i} {t} & = &
\frac {\sum_j G^{i j} x_j} {t^{b \; (2 q - 1)}} \; \; . 
\label{1e3} 
\end{eqnarray}
Equation (\ref{1e3}) determines $b$ in terms of $q \;$ and
relates $\alpha^i$ and $x^i \;$, equivalently $\alpha^i$ and
$c^i \;$ :
\begin{equation}\label{bq}
b \; = \; \frac {1} {2 q - 1} \; \; , \; \; \;
(\gamma \lambda_{qm}) \; \alpha^i \; = \; \sum_j G^{i j} x_j
\; \; . 
\end{equation}
We now analyse equations (\ref{1e4b}) -- (\ref{1e2c}). Note that
$1 + b = 2 b q \;$.


\vspace{4ex} 

\centerline{ $\mathbf \left( \frac {1} {t^{2 b q}} \right) \;
\gg \; \left( \frac {1} {t^{(1 + w) \; \alpha}} \right) \; \; \;
\longrightarrow \; \; $ \bf Asymptotic Anisotropy}

\vspace{2ex} 

Consider first the case where $\left( \frac {1} {t^{2 b q}}
\right) \; \gg \; \left( \frac {1} {t^{(1 + w) \; \alpha}}
\right) \;$ in the asymptotic limits. Then, since $1 + b = 2 b q
\;$, the right hand sides of both the equations (\ref{1e1}) and
(\ref{1e2c}) can be neglected and, hence, the resulting solution
is equivalent to the vacuum solutions obtained earlier. Thus,
one now obtains 
\begin{equation}\label{lggr}
\alpha - b \; = \; \sum_{i j} G_{i j} (c^i)^q (c^j)^q
\; = \; 0   \; \; , 
\end{equation} 
see equations (\ref{w=e1}) with $\rho_0 = 0$ and
(\ref{alphaialpha}). Since $\alpha - b = 0 \;$, equation
(\ref{1e2c}) does not impose any further restriction on $c^i
\;$. Also, with $\alpha = b \;$, equation (\ref{1e4b}) simply
determines the constants $\mu^{i j} \;$ and does not restrict
$(c^i - c^j) \;$. This means that $c^i \;$, and hence $m^i$ and
$\lambda^i_t \;$, are generically different for different $i$
and, therefore, the evolution is anisotropic. The volume factor
$e^\Lambda$ is now given by
\begin{equation}\label{eLaniso}
e^\Lambda \; \propto \; t^\alpha
\; \propto \; t^{ \frac {1} {2 q - 1}} 
\end{equation}
and, depending on the value of $q \;$, it may vanish or diverge
asymptotically.


\begin{itemize}

\item 

Note that the condition $\left( \frac {1} {t^{2 b q}} \right) \;
\gg \; \left( \frac {1} {t^{(1 + w) \; \alpha}} \right) \;$ and
the consequent anisotropic evolution are possible in the
asymptotic limit $t \to 0 \;$ only if $2 b q > (1 + w) \alpha
\;$, and possible in the asymptotic limit $t \to \infty \;$ only
if $2 b q < (1 + w) \alpha \;$. Also, note that $\alpha = b =
\frac {1} {2 q - 1} \;$ and we have assumed that $1 + w > 0 \;$.

\item

Thus, anisotropic evolution is possible in the limit $t \to 0
\;$ only if $2 b q > (1 + w) b \;$. If $2 q > 1$ then $b > 0$
and this inequality implies that $w < 2 q - 1 \;$. If $0 < 2 q <
1$ then $b < 0$ and this inequality implies that $w > 2 q - 1
\;$.

\item

Similarly, anisotropic evolution is possible in the limit $t \to
\infty \;$ only if $2 b q < (1 + w) b \;$. If $2 q > 1$ then $b
> 0$ and this inequality implies that $w > 2 q - 1 \;$.  If $0 <
2 q < 1$ then $b < 0$ and this inequality implies that $w < 2 q
- 1 \;$.

\item

If $q < 0$ then $b < 0 \; , \; b q > 0$ and, hence, $2 b q > (1
+ w) \alpha \;$ always. This means that, when $q < 0 \;$,
anisotropic evolution is possible only in the asymptotic limit
$t \to 0 \;$.

\item

Note that equations (\ref{lggr}) are the analogs of Kasner's
solutions in Einstein's theory. Indeed, for $q = 1 \;$,
equations (\ref{bq}) give $b = 1 \;$ and $(\gamma \lambda_{qm})
\; \alpha^i = c^i \;$; and, equation (\ref{lggr}) then gives
\[
\alpha - 1 \; = \; 0 \; \; , \; \; \;
\alpha^2 - \sum_i (\alpha^i)^2 \; = \; 0 \; \; .
\]
Also, since $2 q - 1 = 1 \;$ now, one has the familar result
that the evolution is anisotropic in the limit $t \to 0$ if $w <
1 \;$, and it is anisotropic in the limit $t \to \infty$ if $w >
1 \;$.

\end{itemize}


\vspace{4ex} 

\centerline{ $\mathbf \left( \frac {1} {t^{2 b q}} \right) \; =
\; \left( \frac {1} {t^{(1 + w) \; \alpha}} \right) \; \; \;
\longrightarrow \; \; $ \bf Asymptotic Isotropy}

\vspace{2ex} 

Consider next the case where $\left( \frac {1} {t^{2 b q}}
\right) \; = \; \left( \frac {1} {t^{(1 + w) \; \alpha}} \right)
\;$ in the asymptotic limits. Since $1 + b = 2 b q \;$, the
right hand sides of both the equations (\ref{1e1}) and
(\ref{1e2c}) are now comparable. One then obtains $2 b q = (1 +
w) \; \alpha \;$ and
\begin{equation}\label{l=r}
\alpha \; = \; \frac {2 q} {(2 q - 1) \; (1 + w)}
\; \; , \; \; \;
\alpha - b \; = \; \frac {2 q - 1 - w} {(2 q - 1) \; (1 + w)}
\; \; . 
\end{equation}
Note that this value of $\alpha$ follows in the isotropic case
also, see equation (\ref{qam}). Equation (\ref{1e1}) gives
\begin{equation}\label{2r=l}
\sum_{i j} G_{i j} (c^i)^q (c^j)^q \; = \; 2 \; \gamma^2
\lambda_{qm}^2 \kappa^2 \; c_\rho \; \; . 
\end{equation}

Consider equation (\ref{1e2c}). If $2 q = 1 + w$ then $\alpha -
b = 0 \;$, equation (\ref{1e2c}) is identically satisfied and,
therefore, imposes no restriction on $c^i \;$. Also, since
$\alpha = b$ now, equation (\ref{1e4b}) simply determines the
constants $\mu^{i j} \;$ and does not restrict $(c^i - c^j)
\;$. This means that $m^i$ are generically different for
different $i$ and, hence, the evolution is anisotropic. This
case where $2 q = 1 + w \;$ has been studied in the earlier part
of this section and explicit solutions have also been presented.

Let $2 q - 1 - w \ne 0 \;$. Then $\alpha - b \ne 0 \;$, these
factors now cancel each other in equation (\ref{1e2c}), and one
obtains
\[
c^i \; = \; \frac {\gamma \lambda_{qm} \kappa^2} {d - 1}
\; (2 q - 1) \; (1 + w) \; c_\rho \; \; .
\]
Thus $c^i \;$, and hence $m^i$ and $\lambda^i_t \;$, are same
for all $i \;$. Then $\alpha^i = \frac {\alpha} {d} \;$, see
equation (\ref{qam}) also. This means that the evolution is
isotropic. Now, this conclusion can be consistent with equation
(\ref{1e4b}) only if $ \left( \frac {1} {t^b} \right) \; \gg \;
\left( \frac {1} {t^\alpha} \right) \;$ in the asymptotic
limits. Then, the right hand side of equation (\ref{1e4b}) can
be negelcted which then implies that $c^i - c^j = 0 \;$ for all
$i$ and $j$ and, hence, that the evolution is isotropic. The
volume factor $e^\Lambda$ is now given by
\begin{equation}\label{eLiso}
e^\Lambda \; \propto \; t^\alpha
\; \propto \; t^{ \frac {2 q} {(2 q - 1) \; (1 + w)}} 
\end{equation}
and, depending on the value of $q \;$, it may vanish or diverge
asymptotically.


\begin{itemize}

\item 

Note that the condition $\left( \frac {1} {t^b} \right) \; \gg
\; \left( \frac {1} {t^\alpha} \right) \;$ and the consequent
isotropic evolution are possible in the asymptotic limit $t \to
0 \;$ only if $b > \alpha \;$, and possible in the asymptotic
limit $t \to \infty \;$ only if $b < \alpha \;$. Also, note that
$b = \frac {1} {2 q - 1} \;$ and $\alpha = \frac {2 b q} {1 + w}
\;$ and we have assumed that $1 + w > 0 \;$.

\item 

Thus, isotropic evolution is possible in the limit $t \to 0 \;$
only if $b > \frac {2 b q} {1 + w} \;$. If $2 q > 1$ then $b >
0$ and this inequality implies that $w > 2 q - 1 \;$. If $0 < 2
q < 1$ then $b < 0$ and this inequality implies that $w < 2 q -
1 \;$.

\item

Similarly, isotropic evolution is possible in the limit $t \to
\infty \;$ only if $b < \frac {2 b q} {1 + w} \;$. If $2 q > 1$
then $b > 0$ and this inequality implies that $w < 2 q - 1 \;$.
If $0 < 2 q < 1$ then $b < 0$ and this inequality implies that
$w > 2 q - 1 \;$.

\item

If $q < 0 \;$ then $b < 0 \; , \; \alpha > 0 \;$ and, hence, $b
< \alpha \;$ always.  This means that, when $q < 0 \;$,
isotropic evolution is possible only in the asymptotic limit $t
\to \infty \;$.

\end{itemize}


\vspace{4ex}

\begin{center}

{\bf 8. Conclusion} 

\end{center}

\vspace{2ex}

We now present a brief summary and conclude by mentioning
several issues for further studies. In this paper, we considered
the LQC -- inspired models which generalise the effective
equations of LQC to $(d + 1)$ dimensions and the function $sin
\; x$ to an arbitrary function $f(x) \;$, see equations
(\ref{hgrav}) and (\ref{genpsi}). Then, assuming that $p = w \;
\rho$ and $(1 + w) > 0 \;$, we studied a variety of $(d + 1)$
dimensional cosmological evolutions in these models
corresponding to a variety of possible behaviours of the
function $f(x) \;$. We found explicit solutions for the
isotropic cases when $f(x) = sin \; x \; , \; x^q \;$, and $e^x
\;$. For these functions, we also found the potential
$V(\sigma)$ for a minimally coupled scalar field $\sigma$ which
may give rise to the equation of state $p = w \; \rho \;$. We
found anisotropic Kasner--type solutions when $f(x) = x^q \;$
and $w = 2 q - 1 \;$.

Together with the shift and the scaling symmetries of the
effective equations, we then used the explicit solutions to
describe the asymptotic evolutions in other examples of $f(x)
\;$ : examples where $f(x) \to x$ in the limit $x \to 0 \;$ and
$f(x) \propto (x_r - x)^q $ in the limit $x \to x_r \;$ or, in
the limit $x \to \infty \;$, the function $f(x) \propto x^q \;$,
or $f(x) \to (const) \;$, or $f(x) \propto e^{- b \; x} \;$.
Such asymptotic behaviours are quite natural and, therefore,
they may apply to a wide class of functions $f(x) \;$.

We find that, depending on $f(x) \;$ in the LQC -- inspired
models, a variety of cosmological evolutions are possible,
singular as well as non singular. Even in the cases where there
is a bounce and no singularities, the asymptotic evolutions are
generically different from that in Einstein's theory. We also
found an intriguing and fascinating result that, for $f(x) =
\sqrt{x} \;$, the evolution is non singular and $a(t)$ grows
exponentially at a rate set by $\lambda_{qm}$, the quantum
parameter related to the area quantum. But its significance, if
any, is not clear to us at present.

We now conclude by mentioning several issues for further
studies. It is worthwhile to understand whether, how, and in
what fundamental theories, the effective equations of the LQC --
inspired models and, in particular, the function $f(x)$ may
arise. One may also study whether similar effective equations
can be constructed, even if only empirically, and applied to
black hole singularities. It will indeed be interesting if black
hole singularities may also be resolved in a variety of ways
depending on some empirical function(s) in such models. The
present LQC -- inspired models may also be applied in the
context of M theory cosmology where, due to U duality symmetries
and appropriate intersecting brane configurations, $(10 + 1)$
dimensional early universe evolves to a $(3 + 1)$ dimensional
universe, with the remaining seven directions remaining constant
in size \cite{k06, mathur, k07, k10}.

Note that one way to adapt our LQC--inspired model to study
black hole singularities would be to construct the covarint
version of the effective equations (\ref{e1}) -- (\ref{e3})
which would generalise Einstien's equations (\ref{rab}). For
various proposals for covariantising the effective equations in
LQC, see \cite{o08} -- \cite{o10}. Such a covariant formulation
is likely to have other applications too. For example, it may be
used to study the evolutions of cosmological perturbations
across a bounce. 

Making progress on any of these issues seems difficult but it
also seems worthwhile to pursue them.


\vspace{4ex}

{\bf Acknowledgement:} 
We thank G. Date for helpful comments. 


\end{document}